\documentclass[journal]{IEEEtran}
\usepackage{amssymb,amsmath,theorem}
\usepackage{array}
\usepackage{graphicx}
\usepackage{epsfig}
\usepackage{amsfonts}
\usepackage{bm}
\usepackage{arydshln}
\usepackage{subfig}
\usepackage{algorithm}
\usepackage{algorithmic}
\usepackage{multirow} 
\usepackage{stfloats}

\newcolumntype{C}[1]{>{\centering}p{#1}}
\usepackage{rotating,setspace,latexsym,amsmath,epsf,amssymb,bm,color}
\usepackage{cite}
\usepackage{booktabs}

\IEEEoverridecommandlockouts

\ifCLASSINFOpdf
\else
\fi

\begin{document}
%
\title{Interacting Object-Enabled Clustering and Characterization of Distributed MIMO Channels}

%

\author{Yingjie~Xu,~\IEEEmembership{Student member,~IEEE},
        Michiel~Sandra,~\IEEEmembership{Student member,~IEEE},
        Xuesong~Cai,~\IEEEmembership{Senior member,~IEEE},
        Sara~Willhammar,~\IEEEmembership{Member,~IEEE}, and~Fredrik~Tufvesson,~\IEEEmembership{Fellow,~IEEE}
\thanks{This work was supported by the Swedish strategic research area ELLIIT, by NextG2Com funded by the VINNOVA program for Advanced Digitalisation with grant number 2023-00541, and by REINDEER which has received funding from the European Union’s Horizon 2020 research and innovation program under grant agreement No. 101013425. \emph{(Corresponding author: Yingjie Xu.)}

Y.~Xu, M.~Sandra, S.~Willhammar, and F.~Tufvesson are with the Department of Electrical and Information Technology, Lund University, Lund, Sweden (e-mail: \{yingjie.xu, michiel.sandra, sara.willhammar, fredrik.tufvesson\}@eit.lth.se).

X. Cai is with the School of Electronics, Peking University,
Beijing, 100871, China (email: xuesong.cai@pku.edu.cn).}}

%
%

\markboth{Journal of \LaTeX\ Class Files,~Vol.~xx, No.~x, January~2025}%
{Shell \MakeLowercase{\textit{et al.}}: Bare Demo of IEEEtran.cls for IEEE Journals}
%



\maketitle

\begin{abstract}
Distributed multiple-input multiple-output (MIMO), also known as cell-free massive MIMO, emerges as a promising technology for sixth-generation (6G) systems to support uniform coverage and reliable communication.
For the design and optimization of such systems, measurement-based investigations of real-world distributed MIMO channels are essential. In this paper, we present an indoor channel measurement campaign, featuring eight distributed antenna arrays with 128 elements in total. Multi-link channels are measured at 50~positions along a 12-meter user route. A clustering algorithm enabled by interacting objects is proposed to identify clusters in the measured channels.  
The algorithm jointly clusters the multipath components for all links, effectively capturing the dynamic contributions of common clusters to different links. In addition, a Kalman filter-based tracking framework is introduced for cluster prediction, tracking, and updating along the user movement. Using the clustering and tracking results, cluster-level characterization of the measured channels is performed. First, the number of clusters and their visibility at both link ends are analyzed. Next, a maximum-likelihood estimator is utilized to determine the entire cluster visibility region length. 
Finally, key cluster-level properties, including the common cluster ratio, cluster power, shadowing, spread, among others, are statistically investigated. The results provide valuable insights into cluster behavior in typical multi-link channels, necessary for accurate modeling of distributed MIMO channels.     

\end{abstract}


%
\IEEEpeerreviewmaketitle

\section{Introduction}
%
%
%
%
\IEEEPARstart{M}{assive} multiple-input multiple-output (MIMO) has become a pillar technology for fifth-generation (5G) networks to improve spectral and power efficiency~\cite{E. G. Larsson 2014,E. BjÃ¶rnson 2017}. To meet the ever-growing communication demands, new paradigms of massive MIMO tailored for sixth-generation (6G) networks are actively being explored, with distributed (massive) MIMO emerging as a key candidate~\cite{U. Madhow 2014,V. Savic 2015}.
In distributed MIMO systems, the base station (BS) antennas are deployed over a wide geographical area, either in co-located groups (semi-distributed) or as individual elements (fully distributed). When integrated with cell-free networks~\cite{H. Q. Ngo 2017}, distributed MIMO systems aim to provide more uniform coverage and more reliable communication~\cite{H. Q. Ngo 2017,O. T. Demir 2021,H. A. Ammar 2022}. 

An ongoing research direction for distributed MIMO is the characterization of the underlying propagation channels; this includes performing channel measurements to explore key channel properties. 
The study in~\cite{F. Rusek2013}, theoretically demonstrated that with the increase of the number of antennas, the user channels tend to become orthogonal, leading to favorable propagation and channel hardening effects. These properties were empirically observed in~\cite{D. Lschenbrand 2022} through an urban channel measurement campaign involving 32~single antennas that were fully distributed over a 46.5~m range. Distributed MIMO channels supported by UAVs were investigated in~\cite{Choi T. 2022,Choi T. 2021}, where multi-link channels were measured at 1,200 ‘virtual’ access points (APs) distributed over a 240~m range, focusing on analysis of channel gain and signal-to-noise ratio (SNR). An extensive measurement campaign, as reported in~\cite{Y. Zhang2024}, explored more than 20,000 possible micro-cellular AP locations, analyzing essential channel statistical properties such as path loss, shadowing, and delay spread.

The aforementioned measurements are limited in several aspects. Firstly, the studies in~\cite{Choi T. 2022,Choi T. 2021,Y. Zhang2024} focused exclusively on channels with fully distributed single antennas, neglecting semi-distributed antenna setups. In particular, the latter has been shown to achieve rates comparable to the former while requiring fewer APs~\cite{Choi T. 2020}.  
Secondly, these measurements relied on a `virtual' antenna array configuration, i.e., moving arrays to different locations such that multi-link channels can be measured sequentially. With this configuration, it is necessary to keep the environment static during the measurements. However, this not only fails to measure dynamic channels, but is also especially challenging when there are a massive number of channel links, as inherent in distributed MIMO systems. To address these issues, real-time channel measurements with distributed antennas and/or antenna arrays are necessary.
In this context, a channel sounder equipped with twelve fully coherent distributed antennas was reported in~\cite{ C. Nelson2025} for an industrial channel measurement campaign. In addition, outdoor channel measurements employing two distributed uniform planar arrays (UPAs) were performed in~\cite{D. LÃ¶schenbrand 2019-1}. 
Indoor channels with eight distributed UPAs were measured in~\cite{Y. Xu202403,Y. Xu202410}. However, these studies focused mainly on link-level channel characterization, such as path loss, channel hardening, delay and Doppler power spectral density, channel non-stationarity,~etc. Cluster-level channel behaviors, which provide deeper insight into the underlying propagation mechanisms, were not considered.

When the propagation channel is resolved into clusters\footnote{Without loss of generality, a cluster is defined as a group of multipath components (MPCs) here that have similar propagation characteristics and interact with similar scattering objects. This definition is consistent with those of many existing channel models, e.g.,~\cite{L. Liu2012, J. Flordelis2020, 3gpp2024, winnerII,J. Li2019}.}, the dynamic changes of the clusters are the inherent reason for the evolution of channels at the link level. In most cluster-based channel models~\cite{L. Liu2012, J. Flordelis2020, J. Li2019}, each cluster is associated with several visibility regions (VRs), which define the geographical areas where the cluster remains effective. A cluster is considered visible in the channel---meaning that it is contributing to it---only when both the BS and user equipment (UE) are located within its corresponding VRs. 
In distributed MIMO systems, the larger separation between distributed access points (APs) amplifies the dynamic variations of clusters across APs. Consequently, only the APs within a cluster’s VR can `see' the cluster, which is referred to as a common/shared cluster in multi-link channels~\cite{J. Poutanen2012}. Comprehensive characterizing these common clusters is essential, as they play a pivotal role in determining the correlations between multi-link channels~\cite{J. Poutanen2010}, and can have a significant impact on system performance. Furthermore, extracting cluster-level parameters, such as cluster density, power, and spread, is crucial for accurate modeling of distributed MIMO channels. However, in most studies, cluster-level characterization of distributed MIMO channel measurements is not well addressed. In addition to the issue of coherently measuring distributed MIMO channels, this gap is due to the lack of appropriate clustering algorithms that are capable of jointly capturing cluster behavior across multiple links\footnote{In this paper, one (channel) link represents the communication link between a distributed AP and a UE, including channels between all antenna elements at the AP and the UE.}. For example, the clustering algorithms in~\cite{X. Cai2020,N. Czink2007,S. Cheng2015} were initially designed for individual links and do not account for common clusters shared among multiple links. 

In summary, most of the existing distributed MIMO channel measurements were conducted only with fully distributed antenna topologies or virtual antenna arrays. Measurement campaigns, in which distributed antenna arrays are deployed and real-time multi-link channels are accurately measured, are still insufficient in the literature. In addition, previous work has focused mainly on link-level channel properties, while cluster-level characterization of distributed MIMO channels has received limited attention. To address these gaps, this paper presents an indoor distributed MIMO channel measurement campaign conducted at a carrier frequency of 5.7~GHz with a bandwidth of 400~MHz. The main contributions and novelties of this work are summarized as follows.
\begin{enumerate}
        \item An indoor distributed MIMO channel measurement campaign was conducted, where eight distributed UPAs with 128 elements in total were used. Multiple simultaneous links were measured at 50~UE positions along a 12-meter route, allowing statistical investigations of multi-link distributed MIMO channel properties.
        \item 
        A novel interacting object (IO)-enabled clustering algorithm is proposed to identify clusters in the measured channels. A multipath component (MPC) distance (MCD) is defined based on IOs and propagation delays. Using a minimum-MCD-based approach, MPCs across all links at a given snapshot are jointly clustered. The clusters are then predicted, tracked, and updated along the UE route by a Kalman filter. The proposed algorithm effectively identifies common clusters in multi-link channels and captures cluster dynamics as the UE moves.
        \item An in-depth analysis of cluster visibility in the measured channels is conducted. Statistical properties are explored, such as the number of visible clusters and VRs per cluster. A maximum likelihood estimator (MLE) is proposed to infer the complete VR length from the observed VRs. Furthermore, the VR radii at both link ends are estimated, offering valuable references for characterizing common clusters in multi-link channels.   
        \item Key cluster-level properties of the measured multi-link channels are statistically investigated, including the common cluster ratio, cluster power and shadowing, intra-cluster angle and delay spreads, and correlations between cluster-level parameters. These statistical findings are fundamental for accurate modeling of distributed MIMO channels.    
\end{enumerate}

The remainder of the paper is organized as follows. Section~II introduces the measurement setup and data processing. Section~III elaborates on the proposed clustering and tracking algorithm, including a performance validation using the measurement data. In Section~IV, the cluster-level characterization of the measured multi-link channels is presented. Finally, conclusions are drawn in Section~V.

\section{Indoor Distributed MIMO Channel Measurements}
\subsection{Measurement Setup}
\begin{figure}[tb]
	\centering
	\subfloat[]
	{
		\begin{minipage}[tb]{0.35\textwidth}
			\centering
			\includegraphics[width=1\textwidth]{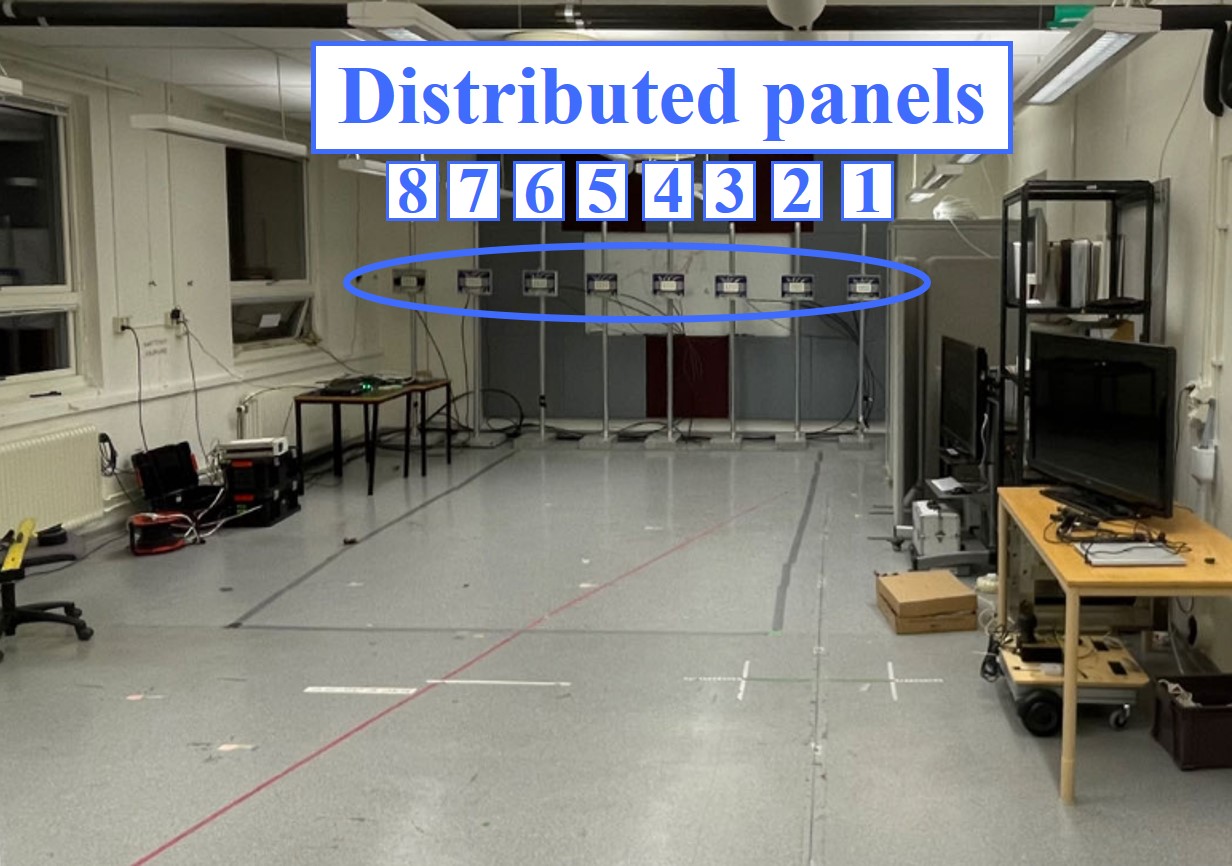}
		\end{minipage}
	}\\
    	\subfloat[]
	{
		\begin{minipage}[tb]{0.225\textwidth}
			\centering
			\includegraphics[width=1\textwidth]{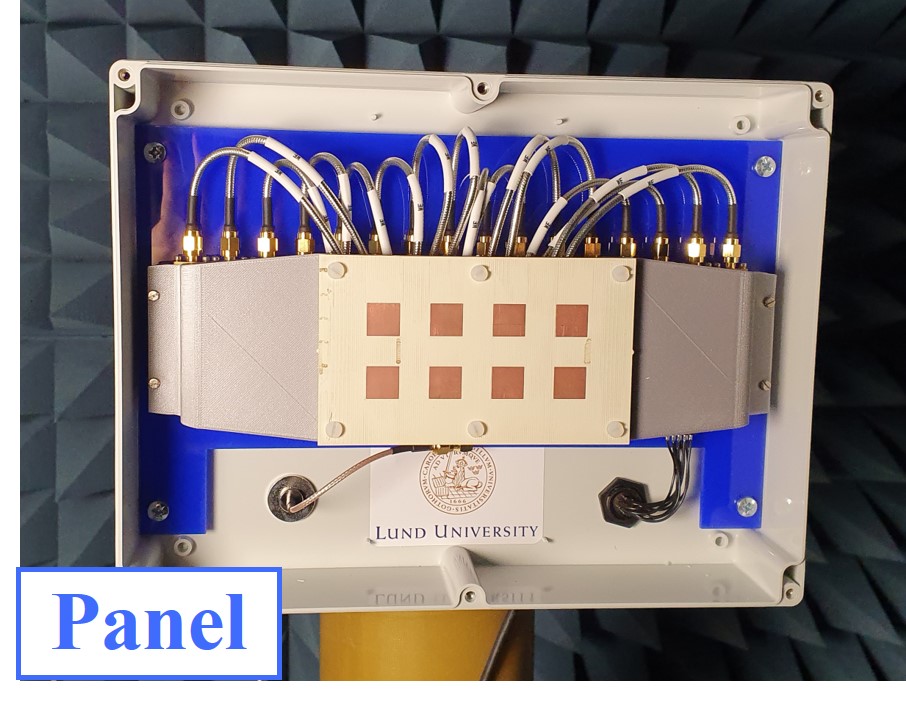}
		\end{minipage}
	}
    \subfloat[]
	{
		\begin{minipage}[tb]{0.25\textwidth}
			\centering
			\includegraphics[width=1\textwidth]{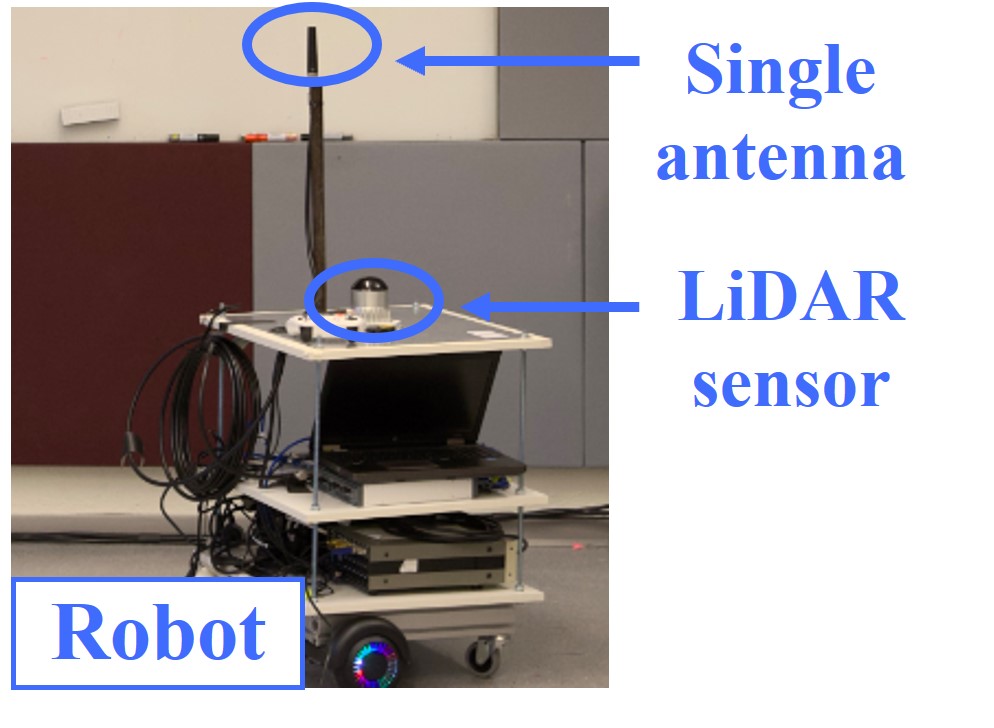}
		\end{minipage}
	}
    
	\caption{Photos of (a) the measurement environment, (b) one of the panels used at the BS side, and (c) the robot at the UE side.}
	\label{Fig_Measurement_environments}
\end{figure}

The channel measurements were carried out in an indoor lab with dimensions 15~m$\times$6~m$\times$2.5~m. The room contained various objects, such as desks, chairs, work station boxes, and screens, as shown in Fig.~\ref{Fig_Measurement_environments}a. These objects contributed to complex propagation mechanisms, including reflection, scattering, diffraction, and signal blocking.

A wideband switched distributed MIMO channel sounder~\cite{M. Sandra2024} was employed for data collection. The sounder integrates multiple NI X410 universal software radio peripherals (USRPs), SP16T radio frequency (RF) switches, and a pair of Rubidium clocks. By combining parallel RF transceiver chains and RF switches, the sounder can measure thousands of antenna combinations within tens of milliseconds, without significantly compromising the dynamic capability. Fig.~\ref{Fig_Measurement_environments}b shows one of the UPAs deployed at the BS side, here referred to as a `panel'. Eight panels were connected to the RF switches of the channel sounder. Designed for operation within the 5–6~GHz frequency range, each panel consists of 2$\times$4 dual-polarized patch antennas (16 ports), with an antenna spacing of 26.7~mm, i.e. approximately half wavelength in this band. On the UE side, a single 3~dBi monopole antenna was used, featuring an omnidirectional radiation pattern and operating within a frequency range of 3.5–8~GHz. The antenna is fixed on a robot that can move via remote control, as shown in Fig.~\ref{Fig_Measurement_environments}c. Rubidium clocks were used to provide a 1~pulse-per-second output signal and a 10~MHz reference clock for synchronizing the BS and the UE.  

\begin{figure}[tb]
	\centering
	{
		\begin{minipage}[tb]{0.48\textwidth}
			\centering
			\includegraphics[width=1\textwidth]{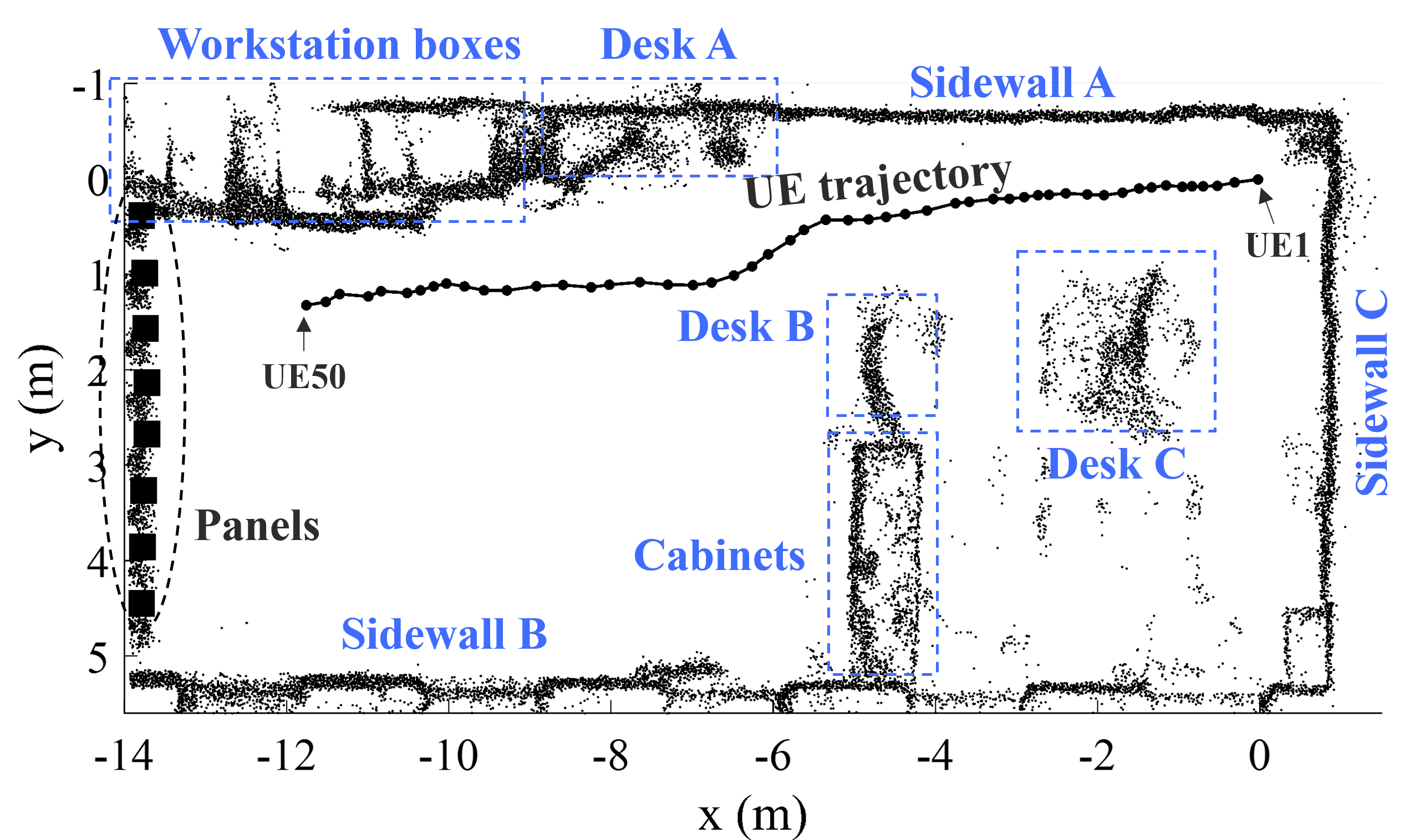}
		\end{minipage}
	}
	\caption{Lidar point cloud of the measurement environment.}
	\label{Fig_Measurement_environment_lidar}
\end{figure}

The channels were measured at a center frequency of 5.7~GHz with a bandwidth of 400~MHz. A Zadoff-Chu sequence of length 1024 was chosen as sounding signal because of its favorable peak-to-average-power ratio and autocorrelation properties. During the measurements, the panels on the BS side were arranged along one side of the room, spaced approximately 60~cm apart. Note that while other antenna topologies may also be promising for distributed MIMO/cell-free massive MIMO systems, the current setup provides an efficient way to continuously capture the dynamic channel behavior across the panels. The robot moved at a constant speed of 0.012~m/s along a trajectory, starting from one end of the room and advancing toward the panels. Its trajectory and positions were accurately tracked by a light detection and ranging (Lidar) sensor, which simultaneously recorded a point cloud of the physical environment, as shown in Fig.~\ref{Fig_Measurement_environment_lidar}. In total, uplink channels involving eight distributed panels were measured along a 12-meter UE route, with channels from 50~distinct UE positions~(snapshots).

\subsection{Measurement Data Processing}
Before measurements are taken, back-to-back calibration is performed to eliminate the response of the measurement system, connectors, and cables. Let $u(t)$ denote the transmit signal used for both the calibration and measurements. Define the responses of the measurement system, connectors, and cables as $G(f)$. After the back-to-back calibration, the received calibration signal is given as $y_c(t)=u(t)*g(t)$,
where $*$ represents the convolution operation, and $g(t)$ is the inverse Fourier transform (IFT) of $G(f)$. During the measurement, the received signal at the BS end is expressed as $y(t)=u(t)*g(t)*h(t).$ Then the CIR $h(t)$ is obtained by
\begin{equation}
	h(t)=\text{IFT}\{H(f)\}=\text{IFT}\{Y(f)/Y_c(f)\},
\end{equation}
 where $Y(f)$ and $Y_c(f)$ are the Fourier transforms of $y(t)$ and $y_c(t)$, respectively.

To extract MPC parameters, the high-resolution space-alternating generalized expectation maximization (SAGE) algorithm~\cite{B. H. Flury 1996, X. Yin 2003, B. H. Flury 1999} is employed. In this algorithm, the generic dual-polarized signal model with the transmitted signal vector $\mathbf{u}(t) $ is in single-input multiple-output channels described as~\cite{X. Yin 2003}  
\begin{equation}
     \begin{split}
      \mathbf y(t) = &\underset{l = 1}{\overset{L}{\sum }}\mathrm{\exp}(j2\pi{v}_{l}t){\mathbf c}_{r}({\mathbf{\Omega}}_{l}){\mathbf A}_{l}\mathbf u(t - {\tau}_{l})+ \sqrt{\frac{{N}_{0}}{2}}\mathbf N(t),
  \end{split} 
\end{equation}
where $L$ and $\mathbf N(t)$ denote the extracted number of MPCs in the channel and the complex white Gaussian noise with a spectral density of $N_0$, respectively. The estimated parameters include the delay ${\tau}_{l}$, angle of arrival (AoA) ${\mathbf{\Omega}}_{l}$, Doppler frequency ${v}_{l}$, and the polarization matrix ${\mathbf A}_{l}$, which consists of the dual-polarized complex amplitudes ${{\bf{A}}_l} = \left[ {{\alpha _{l,{\text{V}}},\alpha _{l,\text{H}}}} \right]^{\text{T}}$, $p_1 \in [1,2]$. For a given ${\mathbf{\Omega}}_{l}$, it is characterized by both the azimuth angle ${\phi}_l$ and the elevation angle ${\theta}_l$ in spherical coordinates, i.e., $\mathbf{\Omega}_l = [\cos \phi_l\sin\theta_l,\sin\phi_l\sin\theta_l,\cos\theta_l]^{\text{T}}$. The antenna array response ${\mathbf c}_{r}({\mathbf{\Omega}}_{l})$ is measured in an anechoic chamber. The SAGE algorithm divides the multi-dimensional estimation into multiple low-dimensional sub-estimations, which efficiently reduces the complexity of the expectation-maximization process. For more details on the SAGE algorithm, see~\cite{X. Yin 2003}. In this study, the algorithm is used independently for each panel. The estimated number of MPCs in the algorithm is set to 200, which is found to provide a good trade-off between extracting sufficient channel information and minimizing computational complexity.

\section{Cluster Identification and Tracking}

\subsection{Interacting object-enabled clustering algorithm}
In this section, an IO-enabled clustering algorithm is proposed to identify clusters in the measured channels. It includes two main steps. First, leveraging the collected Lidar point cloud, a ray-launching method is exploited to explicitly map the MPCs to physical objects in the environment, here referred to as IOs. Next, the MPCs from all links are clustered simultaneously based on their `similarities' in terms of their associated IOs and propagation delays. This allows for investigation of whether MPCs from different links belong to the same clusters, that is, whether they form common clusters across multiple links or not.

Considering a double-directional channel model, each MPC is associated with two IOs: one on the UE side and the other on the BS side. These are referred to as first- and last-hop scatterers\footnote{In this paper, uplink is assumed. For the downlink, the locations of the first- and last-hop scatterers have to be swapped.}. The center location of the IOs for each MPC is determined by finding the intersection of the MPC ray with physical objects in the environment. In this measurement setup, the algorithm only considers the IOs on the BS side. However, the method can be easily extended to channels with an MIMO setup.   
 
For the $l$-th MPC$_{k,l}$ in the panel~$k$-UE link, define a ray $\xi_{k,l}$ starting from the location of panel~$k$, i.e., $\mathbf{P}_k=(x_{p_k},y_{p_k},z_{p_k})$. Given the angle of arrival $\mathbf{\Omega}_{k,l} =\left [ \phi_{k,l}, \theta_{k,l}  \right ] $, the parametric equation of $\xi_{k,l}$ is expressed as
\begin{equation}
\xi_{k,l}: \begin{cases}
x = x_{p_k}+d \cdot \cos \phi_{k,l} \cdot \sin \theta_{k,l}, \\
y = y_{p_k}+d \cdot \sin \phi_{k,l} \cdot \sin \theta_{k,l}, \\
z = z_{p_k}+d \cdot \cos \theta_{k,l},
\end{cases}
\label{MPC_ray}
\end{equation}
where $d$ is a distance parameter along the ray with $0< d\leq \tau _{k,l}\cdot c$, where $\tau _{k,l}$ is the delay of MPC$_{k,l}$ and $c$ represents the speed of light. Let $\mathbf{s}=(x_{s},y_{s},z_{s})$ denote the collected Lidar point cloud. The set of potential intersections of the MPC$_{k,l}$ ray and the physical objects in the environment is obtained by a nearest-neighbor search. Specifically, $\mathbf{\kappa}_{k,l}= \{\mathbf{s}| \mathrm{dist}(\mathbf{s}, \xi_{k,l}) \leq \delta_0\}$,
where the operation $\mathrm{dist}(\cdot)$ calculates the distance between a point and a ray in space, and $\delta_0$ is a distance threshold determined by the density of the point cloud. The intersections in $\mathbf{\kappa}_{k,l}$ are then clustered into potential IOs using the DBSCAN algorithm~\cite{M. Ester1996}, which effectively takes into account the density of the dataset. Note that there could be multiple potential IOs; the real IO is determined as the one that $\xi_{k,l}$ hits first. The center location of the real IO can be obtained by
\begin{equation}
\mathbf{o}_{k,l} = \min_{\mathbf{o}_{k,l,j}} \left \|\mathbf{o}_{k,l,j}-\mathbf{P}_k \right \|_F,  
\label{find_IO}
\end{equation}
where $\mathbf{o}_{k,l,j}$ is the center location of $j$-th potential IO as clustered from $\mathbf{\kappa}_{k,l}$, and $\left \| \cdot  \right \|_F $ is the Frobenius norm.
\begin{algorithm}[t]
	\renewcommand{\algorithmicrequire}{\textbf{Input:}}
	\renewcommand{\algorithmicensure}{\textbf{Output:}}
\caption{The Proposed IO-Enabled Clustering Algorithm}
\label{algorithm1}
\begin{algorithmic}[1]
\REQUIRE Estimated parameters of MPC$_l$, $l=1,...,L^{(n)}$ at snapshot $n$, point cloud of the environment $\mathbf{s}$, locations $\mathbf{P}_k$ of panel $k$, $k=1,...,K$
\ENSURE Resulted cluster index $c=1,...,C^{(n)}$, cluster centroid $\Theta_c^{(n)}$, clustered MPC sets $\eta_c^{(n)}$
\STATE Initialize MPC path index $l=1$
\WHILE{$l \le L^{(n)}$}
\STATE Find the IO center $\mathbf{o}_{k,l}$ of MPC$_{k,l}$, $k=1,...,K$ by~(\ref{MPC_ray}) and (\ref{find_IO}) \\
\STATE Calculate the partial delay $\tau_{{\rm{P}},k,l}$ of MPC$_{k,l}$ by~(\ref{partial_delay})
\STATE $l=l+1$\\
\ENDWHILE
\STATE Initialize set $\eta= \{1,...,L^{(n)}\} $ containing all MPC indexes at snapshot $n$. Initialize cluster index $c=1$
\WHILE {$\eta\ne \emptyset$}
\STATE Find the MPC with the highest power $l_{o}= \arg\max_{l'\in\eta}\left |\alpha_{l'} \right | ^2$ 
\STATE Regard MPC${_{l_o}}$ as the reference point of cluster $c$
\STATE Initialize MPC set $\eta_c^{(n)}=\emptyset$
\FOR{$l\in \eta$}
\STATE Calculate MCD$_{l,l_0}$ according to (\ref{MCD_IO}), (\ref{MCD_tau}), and (\ref{MCD_all})
\IF{MCD$_{l,l_0} \leq \delta_{\rm{MCD}}$}
\STATE $\eta_c^{(n)}=\eta_c^{(n)}\cup \{l\}$, $\eta=\eta \setminus \{l\}$
\ENDIF
\ENDFOR
\STATE $c=c+1$
\ENDWHILE
\STATE Update cluster index $C_{(n)}=c-1$
\STATE Calculate cluster centroid $\Theta_c^{(n)}, c=1,...,C_{(n)}$ by~(\ref{IO_cluster}) and~(\ref{partial_delay_cluster})
\WHILE{$\exists c, \Theta_c^{(n)}$ changes over iterations}
\STATE Let $\eta= \{1,...,L^{(n)}\} $, and initialize MPC set $\eta_c^{(n)}=\emptyset, c=1,...,C_{(n)}$ 
\WHILE {$\eta\ne \emptyset$}
\FOR{$l\in \eta$}
\STATE Find the cluster $c'$ whose centroid has the minimum MCD with MPC$_l$
\STATE Update $\eta_{c'}^{(n)}=\eta_{c'}^{(n)}\cup \{l\}$, $\eta=\eta \setminus \{l\}$

\ENDFOR
\ENDWHILE
\STATE Calculate cluster centroid $\Theta_c^{(n)}, c=1,...,C_{(n)}$ by~(\ref{IO_cluster})  
\ENDWHILE
\end{algorithmic}
\end{algorithm}
The minimum MCD-based method is exploited for MPC clustering because of its advantage of not requiring prior knowledge about the number, shape, or distribution of the clusters. The MCD measures the multipath separation in the delay and angle domains~\cite{X. Cai2020}. However, its calculation is based on the local coordinate system of individual links, indicating that it cannot be directly applied to multi-link channels. To address this limitation, a modified minimum MCD-based method is proposed. Unlike the definition in~\cite{X. Cai2020}, one component of the MCD is now defined as the differences between MPC $i$ in the panel~$k_1$-UE link and MPC $j$ in the panel~$k_2$-UE link in terms of their corresponding IO locations, i.e.
\begin{equation}
{\rm{MCD}}_{{\rm{IO}},ij}=\left \| \mathbf{o}_{k_1,i}-\mathbf{o}_{k_2,j} \right \| _F.
\label{MCD_IO}
\end{equation}
When $k_1=k_2$, MPC $i$ and $j$ are within the same link. Since IOs are defined in a global coordinate system, the modified MCD in (\ref{MCD_IO}) allows us to quantify the differences between two MPCs even if they are on different links. Next, the separation between two MPCs is further measured in terms of their experienced propagation, which can be characterized by their propagation delay. Note that even if two MPCs in different links interact with similar scatterers when propagating, their propagation paths between the last-hop scatterers and the panels may differ significantly. This is due to the separation between the panels, which makes it invalid to define MCD by their total propagation delay, as used in~\cite{X. Cai2020}. Instead, we focus on the propagation of MPC$_{k,i}$ before the last-hop scatterer, and denote this as a partial propagation delay $\tau_{{\rm{P}},k,i}$, which is calculated by 
\begin{equation}
\tau_{{\rm{P}},k,i}=\cfrac{\tau_{k,i}\cdot c-\left \|\mathbf{o}_{k,i}-\mathbf{P}_k \right \|_F }{c}.  
\label{partial_delay}
\end{equation}
\newcounter{TempEqCnt}  
\setcounter{TempEqCnt}{\value{equation}} 
\setcounter{equation}{12}  
\begin{figure*}[hb]
\hrulefill
\begin{equation}
	\begin{split}
 		\mathcal{G}_c(\Theta_{c'}^{(n)} | \tilde{\Upsilon} _c^{(n|n-1)}, \mathbf{C}_c) = \frac{1}{(2 \pi)^{3/2} |\mathbf{C}_c|^{1/2}} \cdot \exp \left( -\frac{1}{2} (\Theta_{c'}^{(n)} - \tilde{\Upsilon} _c^{(n|n-1)})^T
\cdot \mathbf{C}_c^{-1} (\Theta_{c'}^{(n)} - \tilde{\Upsilon} _c^{(n|n-1)}) \right)
	\end{split}
	\label{eq:closeness function}
\end{equation} 
\end{figure*}
\setcounter{equation}{\value{TempEqCnt}}
Then the MCD in the delay domain between MPC$_{k_1,i}$ and MPC$_{k_2,j}$ is given by~\cite{X. Cai2020}
\begin{equation}
{\rm{MCD}}_{\tau,ij}=\zeta \cdot\cfrac{\left |\tau_{{\rm{P}},k_1,i}-\tau_{{\rm{P}},k_2,i}  \right | }{\Delta \tau_{{\rm{P}},max}} \cdot\cfrac{\tau_{{\rm{P,std}}}}{\Delta \tau_{{\rm{P}},max}}, 
\label{MCD_tau}
\end{equation}
where $\Delta \tau_{{\rm{P}},max}=\max_{k_1,k_2,i,j} \left |\tau_{{\rm{P}},k_1,i}-\tau_{{\rm{P}},k_2,j}  \right |$, and $\tau_{{\rm{P,std}}}$ is the standard deviation of the partial delays of MPCs across all the links. A delay scaling factor $\zeta$ is introduced to give the delay domain a weight, which is useful when clustering real-world data~\cite{X. Cai2020}. Finally, the overall MCD between two MPCs is given by considering both ${\rm{MCD}}_{{\rm{IO}},ij}$ and ${\rm{MCD}}_{\tau,ij}$, namely,
\begin{equation}
{\rm{MCD}}_{ij}=\sqrt{{\rm{MCD}}^2_{{\rm{IO}},ij}+{\rm{MCD}}^2_{\tau,ij}}. 
\label{MCD_all}
\end{equation}

 The proposed clustering algorithm starts by initializing the clusters through an iterative assignment of MPCs that are sufficiently close to a selected initial reference point. In this case, the initial reference point is chosen as the MPC with the highest power among the remaining unassigned MPCs. The term `sufficiently close' refers to the condition where the MCD between a candidate MPC and the reference point is smaller than a predefined threshold $\delta_{\rm{MCD}}$. After the initial assignment, the centroid of a cluster~$c$ is defined by its IO center and partial delay, represented as $\Theta _c=[\mathbf{o}_{c},\tau_{{\rm{P}},c}]^{\rm{T}}$, with
 \begin{equation}
\mathbf{o}_{c}=\cfrac{\sum_{l\in\eta_c}\left | \mathbf{A}_l  \right |^2\cdot \mathbf{o}_{k,l}}{\sum_{l\in\eta_c}\left | \mathbf{A}_l  \right |^2} 
\label{IO_cluster}
\end{equation}
  \begin{equation}
\tau_{{\rm{P}},c}=\cfrac{\sum_{l\in\eta_c}\left | \mathbf{A}_l  \right |^2\cdot \tau_{{\rm{P}},k,l}}{\sum_{l\in\eta_c}\left | \mathbf{A}_l  \right |^2}, 
\label{partial_delay_cluster}
\end{equation}
where $\eta_c$ denotes the set of MPCs assigned to cluster~$c$. Subsequently, the MPCs are iteratively assigned to the cluster with the minimum MCD. At the same time, the centroid of the cluster is updated accordingly. The algorithm continues until the assignment of MPCs stabilizes and no further changes occur between iterations. Details of the proposed algorithm is provided in Algorithm~1. The resulting $\Theta_{c}$ at each snapshot will be used as the observed data for cluster tracking over snapshots, as introduced in the following section.

\subsection{Kalman filter-based cluster tracking framework}
\begin{figure}[t!]	\centerline{\includegraphics[width=0.38\textwidth]{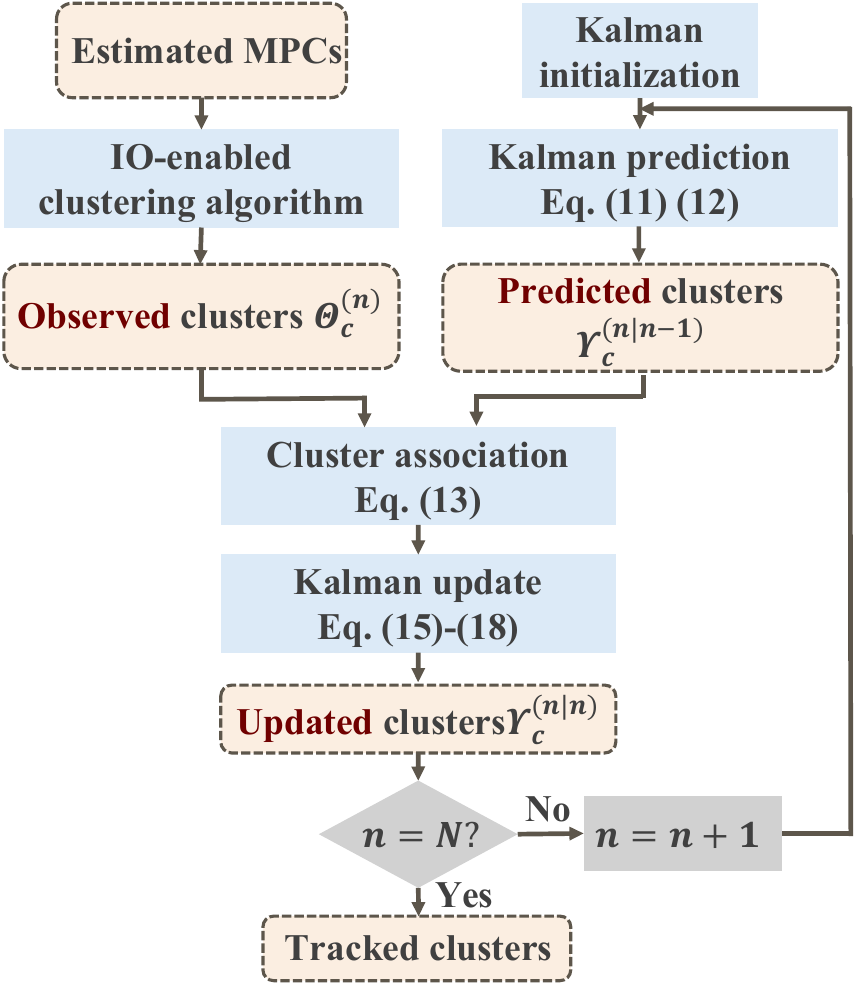}}
	\caption{Flow chart of the Kalman filter-based cluster tracking framework.}
	\label{Fig_flow_chart}
\end{figure}

To evaluate the time-varying behavior of the clusters, tracking their evolution over snapshots is essential. Several cluster tracking algorithms have been proposed, including minimum distance-based tracking~\cite{X. Cai2020}, trajectory-based tracking~\cite{C. Huang2020}, and Kalman filter-based tracking~\cite{N. Czink2007}. However, most of these methods only track clusters between consecutive snapshots. In practice, due to potential obstructions or limited performance of the channel parameter estimation algorithm, the same clusters may disappear for some snapshots and then reappear later. In this section, considering the possible reappearance of clusters, the Kalman filter-based tracking framework from~\cite{N. Czink2007} is modified to predict, track, and update clusters over snapshots. A flow chart of the framework is given in Fig.~\ref{Fig_flow_chart}, with the details introduced below.
  \begin{figure}[tb!]
	\centerline{\includegraphics[width=0.48\textwidth]{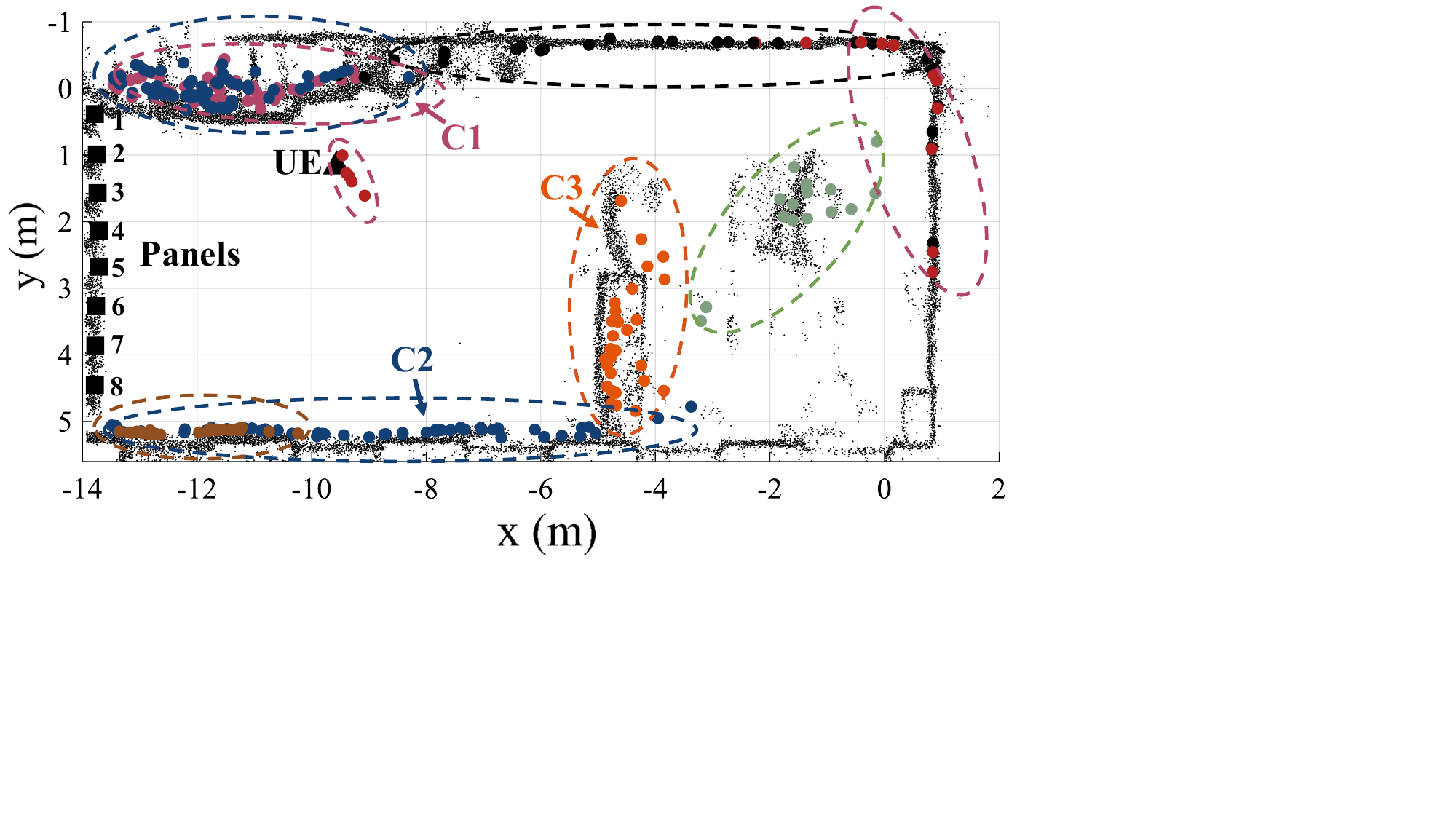}}
	\caption{Example of cluster identification results visualized on the captured Lidar point cloud.}
	\label{Fig_cluster_results_sn10}
\end{figure}

Assuming linear motion of the cluster centroid, the state model of the cluster centroid $\Upsilon _c^{(n)}=[x_{\mathbf{o}_{c}},\Delta x_{\mathbf{o}_{c}},y_{\mathbf{o}_{c}},\Delta y_{\mathbf{o}_{c}},z_{\mathbf{o}_{c}},\Delta z_{\mathbf{o}_{c}},\tau_{\text{P},c},\Delta \tau_{\text{P},c}]^{\text{T}}$ at snapshot $n$ is given by $\Upsilon _c^{(n)}=\Phi \Upsilon _c^{(n-1)}+\mathbf{w}^{(n)}$, where $\mathbf{w}^{(n)}$ represents the state-noise with covariance matrix $Q$, and the state-transition matrix $\Phi$ is given by $\Phi=\mathbf{I}_4\otimes \begin{bmatrix}
        1  & 1 \\ 0  & 1 \end{bmatrix} $, where $\mathbf{I}_4$ is a four-dimensional identity matrix and $\otimes$ denotes the Kronecker matrix product.
Furthermore, the measurement model for the cluster tracking is given by $\Theta _c^{(n)}=\mathbf{H}\Upsilon _c^{(n)}+\mathbf{v}^{(n)}$, where $\Theta _c^{(n)}$ denotes the observed cluster centroid, which is the output from Algorithm~1, $\mathbf{v}^{(n)}$ is the measurement noise with covariance matrix $\mathbf{R}$, and the measurement matrix $\mathbf{H}$ is given by
$\mathbf{H}=\mathbf{I}_4\otimes \begin{bmatrix}        1   & 0  \end{bmatrix}.$ 
Given the results from the previous snapshot, the predicted cluster centroid $\Upsilon _c^{(n|n-1)}$ and error covariance matrix $\mathbf{M}^{(n|n-1)}$ at snapshot $n$ are followed by~\cite{N. Czink2007}
\begin{equation}
\Upsilon _c^{(n|n-1)}=\Phi \Upsilon _c^{(n-1|n-1)}
\label{KF_predict1}
\end{equation}
\begin{equation}
\mathbf{M}^{(n|n-1)}=\Phi \mathbf{M}^{(n|n-1)}\Phi ^{\rm{T}}+\rm{Q}.
\label{KF_predict2}
\end{equation}
Next, cluster association needs to be performed between the predicted clusters and the observed clusters. A probability-based method~\cite{N. Czink2007} is considered for this purpose. Specifically, for each predicted clusters with centroid $\Upsilon _c^{(n|n-1)}$ at snapshot $n$, select an observed cluster $c'$ with centroid $\Theta_{c'}^{(n)}$ that has the maximum closeness function among them. The closeness function is defined as (\ref{eq:closeness function}) at the bottom of this page, where $\tilde{\Upsilon}_c^{(n|n-1)}=[x^{(n|n-1)}_{\mathbf{o}_{c}},y^{(n|n-1)}_{\mathbf{o}_{c}},z^{(n|n-1)}_{\mathbf{o}_{c}},\tau^{(n|n-1)}_{\text{P},c}]^{\text{T}}$ is the partial state of $\Upsilon _c^{(n|n-1)}$, and $\mathbf{C}_c$ denotes the cluster spread matrix, which is given by
\setcounter{equation}{13}
\begin{equation}
\mathbf{C}_c=\cfrac{\sum_{l\in\eta_c}\left | \alpha_l  \right |^2\cdot (\mathbf{x}_l-\tilde{\Upsilon} _c^{(n|n-1)})(\mathbf{x}_l-\tilde{\Upsilon} _c^{(n|n-1)})^T}{\sum_{l\in\eta_c}\left | \alpha_l  \right |^2}, 
\end{equation}
where $\mathbf{x}_l=[\mathbf{o}_{k,l},\tau_{{\rm{P}},k,l}]^T$. Similarly, for each observed cluster, a predicted cluster is selected based on the maximum value of $\mathcal{G}_c'(\tilde{\Upsilon} _c^{(n|n-1)} |\Theta_{c'}^{(n)},\mathbf{C}_{c'})$. On the one hand, if a predicted cluster and an observed cluster are mutually selected, these two clusters are viewed as tracked clusters. Based on (\ref{KF_predict1}), (\ref{KF_predict2}), and the resulting $\Theta _c^{(n)}$, the updated $\Upsilon _c^{(n|n)}$, Kalman gain $K^{(n)}$, and error covariance matrix $\mathbf{M}^{(n|n)}$ are given by
\begin{equation}
K^{(n)} = \mathbf{M}^{(n|n-1)} \mathbf{H}^T \left( \mathbf{H} \mathbf{M}^{(n|n-1)} \mathbf{H}^T + \mathbf{R} \right)^{-1} 
\label{KF_update1}
\end{equation}
\begin{equation}
\mathbf{M}^{(n|n)} = \left( \mathbf{I} - K^{(n)} \mathbf{H} \right) \mathbf{M}^{(n|n-1)}
\label{KF_update2}
\end{equation}
\begin{equation}
\Upsilon _c^{(n|n)} = \Upsilon _c^{(n|n-1)} + K^{(n)} \left( \Theta _c^{(n)} - \mathbf{H} \Upsilon _c^{(n|n-1)} \right).
\end{equation}
On the other hand, observed clusters that are not mutually associated with any predicted clusters, are marked as `born' clusters. Similarly, predicted clusters that are not associated with any observed clusters are marked as `disappeared' clusters in the current snapshot. However, in contrast to the operations in~\cite{C. Huang2020}, updates for the `disappeared' clusters continue, as described by (\ref{KF_update1}), (\ref{KF_update2}), and
\begin{equation}
\Upsilon _c^{(n|n)} = \Upsilon _c^{(n|n-1)}.
\end{equation}
These updates continue until the cluster has been marked as `disappeared' for a prolonged period, i.e., for more than $n_{th}$ snapshots. At this point, the cluster is considered `dead', and its lifetime is terminated at the final snapshot in which it appeared. Note that $n_{th}$ should be large enough to effectively track clusters that disappear for a certain period but also small enough to prevent incorrect associations between different clusters. Its setting depends on the specific channel conditions, i.e. channel stationarity, and measurement configurations, including snapshot rate and UE mobility. Here we use $n_{th}=5$~snapshots. Using the introduced framework, clusters are tracked over snapshots, allowing studies of cluster behavior over time. 


\subsection{Cluster identification and tracking results}
In this section, the proposed clustering and tracking algorithm is validated with the measurement data. According to the density of the collected point cloud, the threshold $\delta_0$ is here set to 0.5. Furthermore, following the approach suggested in~\cite{X. Cai2020}, a metric based on multiple cluster validity indices (CVIs), is considered to select an appropriate threshold value $\delta_{\rm{MCD}}$. More specifically, $\delta_{\rm{MCD}}$ is chosen as the point when the resulting clusters exhibit both reasonable compactness and separability, which is evaluated through multiple CVIs. In the case here, a suitable $\delta_{\rm{MCD}}$ is determined to be 7, and the delay scaling factor $\zeta$ in the algorithm is set to 1. Note that these parameters may vary depending on the specific measurement environment, for example, in outdoor scenarios, where the separation of objects might be more significant, a larger $\delta_{\rm{MCD}}$ might be more suitable.

Fig.~\ref{Fig_cluster_results_sn10} illustrates an example of cluster identification results for a specific snapshot overlayed on the captured Lidar point cloud. The panels and the UE are shown as black squares and triangles, respectively. Each dot corresponds to the IO center for each MPC of all the links, where dots of the same color indicate that the MPCs are assigned to the same cluster. For visualization purposes, the figure omits MPCs that interact with the ceiling and the ground. An observation is that, as expected, there is consistency between the line-of-sight (LoS) components and the actual UE position. In addition, most clusters are reasonably well connected to physical objects in the environment, such as sidewalls, desks, and workstation boxes. It should be noted that some MPCs, despite having `similar' IO centers, are not assigned to the same cluster---an example being the MPCs with IO centers near the work station boxes. This is due to their different behavior in the delay domain, characterized by their partial delays in (\ref{partial_delay}), thus resulting in a large MCD between them. This phenomenon can be explained by the fact that, although the last-hop scatterers for these MPCs are located close to each other, the propagation paths before the last-hop scatterers differ significantly. Such behavior is reasonable, particularly in indoor environments, where rich scattering typically leads to various propagation mechanisms influencing the signals~\cite{L. Possenti2023}.

\begin{figure}[t]
	\centering
  	\subfloat[]
	{
		\begin{minipage}[tb]{0.40\textwidth}
			\centering
			\includegraphics[width=1\textwidth]{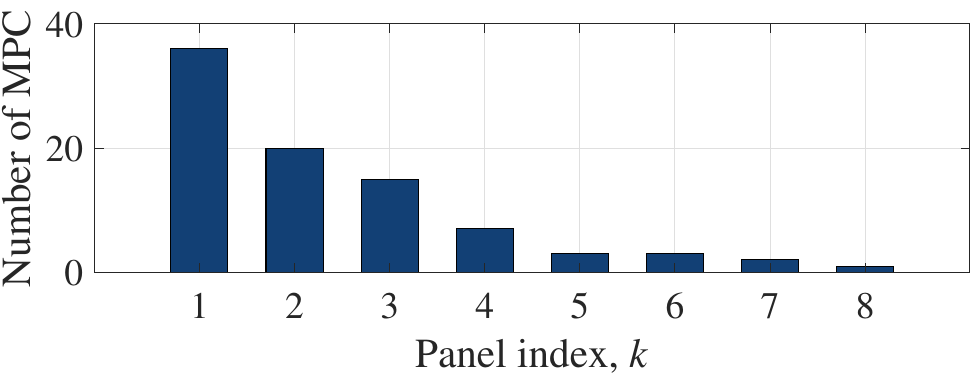}

		\end{minipage}
	}
    \vspace{-0.05cm}
  	\subfloat[]
	{
		\begin{minipage}[tb]{0.40\textwidth}
			\centering
			\includegraphics[width=1\textwidth]{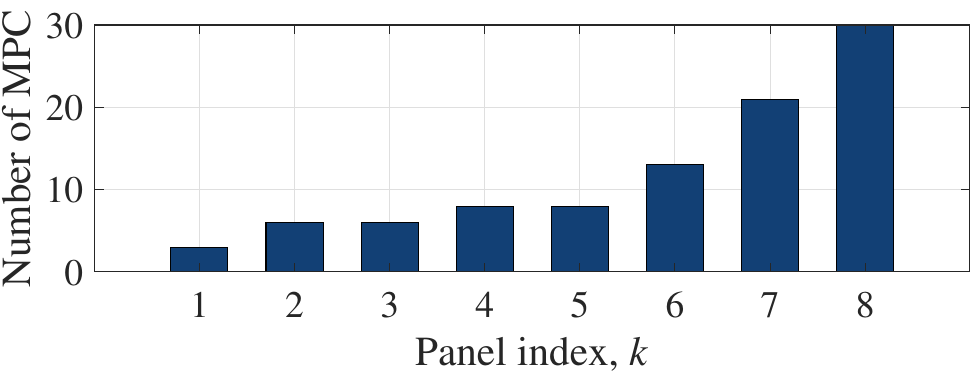}

		\end{minipage}
	}\vspace{-0.05cm} 
  	\subfloat[]
	{
		\begin{minipage}[tb]{0.40\textwidth}
			\centering
			\includegraphics[width=1\textwidth]{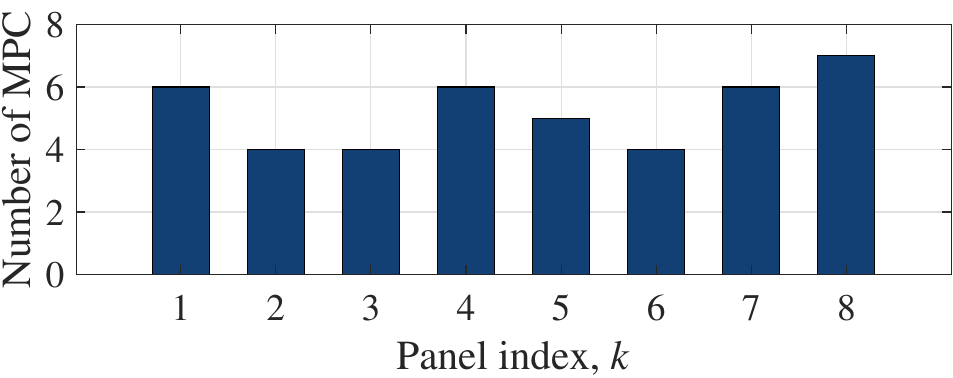}

		\end{minipage}
	}

	\caption{Distribution of MPCs within the example clusters (a) C1, (b) C2, and (c) C3.}
	\label{Fig_example clusters}
\end{figure}

Furthermore, the distribution of MPCs between a panel~$k$ and the UE, where $k=1,...,8$ is further detailed in Fig.~\ref{Fig_example clusters}a--\ref{Fig_example clusters}c for the clusters~C1-C3 (see Fig.~\ref{Fig_cluster_results_sn10}). It can be observed that the number of MPCs that contribute to the different links varies within a cluster, indicating that the influence of a cluster on different links is dynamic. Moreover, within a cluster, links with panels located closer to the cluster centroid tend to have a higher number of MPCs. This trend is expected since panels closer to a cluster are more likely to receive signals emanating from it. For instance, in cluster~C1, stemming from the work station boxes in Fig.~\ref{Fig_cluster_results_sn10}a, most MPCs are found to be from the link to panel~1, which is the panel closest to the cluster. As the distance between the cluster and subsequent panels increases, the number of associated MPCs decreases, ultimately leaving only one MPC from the link to panel~8, the farthest panel. A similar trend can be observed in Fig.~\ref{Fig_example clusters}b, where the majority of MPCs are contributing to the link to panel~8, the nearest panel to cluster~C2, as originating from sidewall~B. In contrast, cluster~C3 is identified as the cabinets and surrounding desks and exhibits a relatively even distribution of MPCs across all links. This is because the distances between this cluster and the panels are approximately equal, leading to similar contributions from the cluster to each link.

\begin{figure}[t]
	\centering
	\subfloat[]
	{
		\begin{minipage}[t]{0.47\textwidth}
			\centering
			\includegraphics[width=1\textwidth]{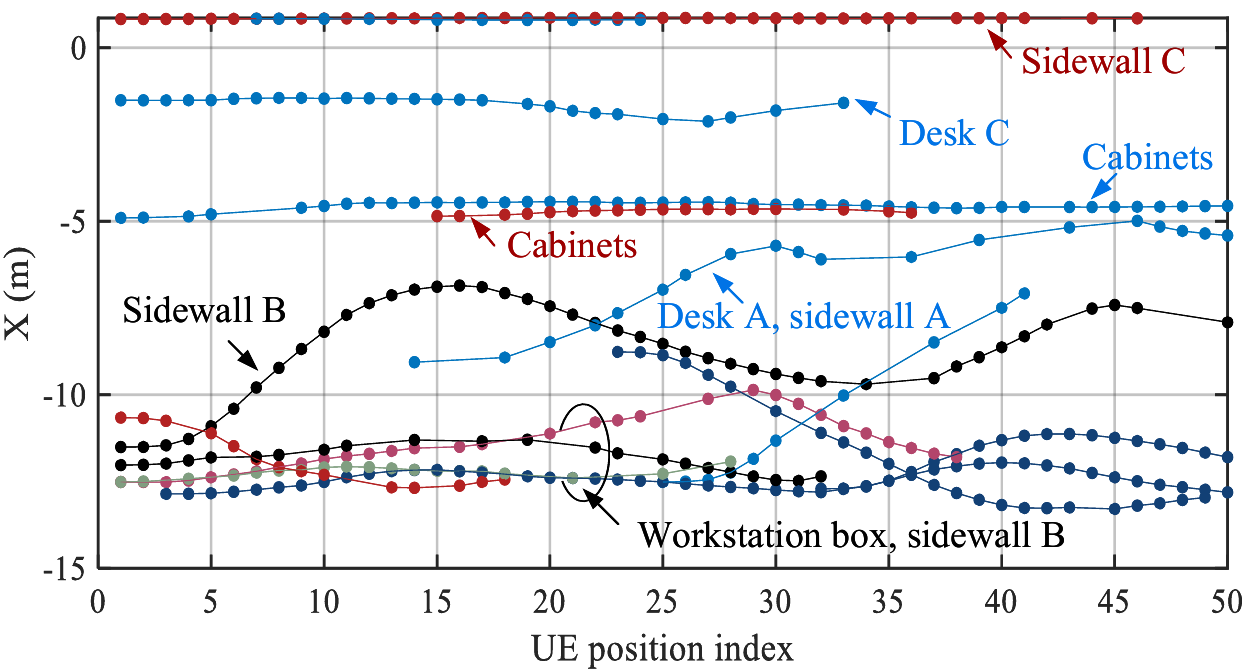}

		\end{minipage}
	}\\
	\subfloat[]
	{
		\begin{minipage}[t]{0.47\textwidth}
			\centering
			\includegraphics[width=1\textwidth]{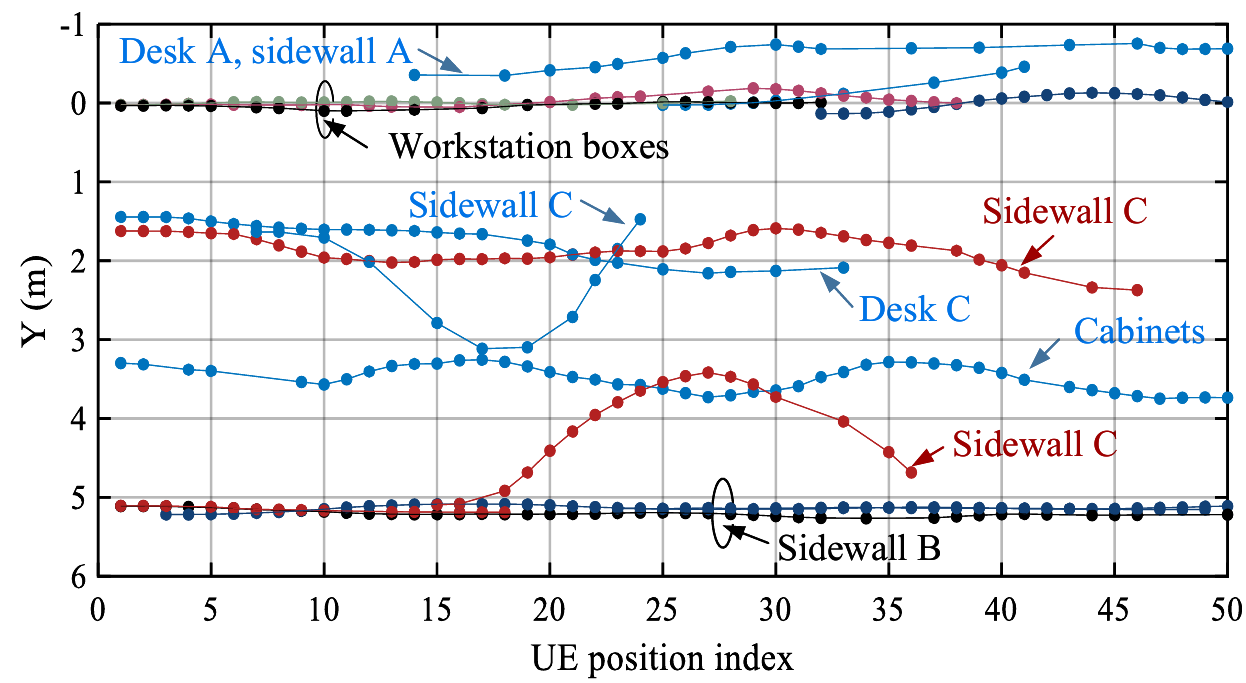}

		\end{minipage}
	}
	\caption{(a) X and (b) Y coordinates of the cluster IO center vs. UE position.}
	\label{Fig_tracking_cluster_xyz}
\end{figure}

To evaluate the performance of the presented cluster tracking framework, the X and Y coordinates of the cluster IO centers, which change with the movement of the UE, are shown in Fig.~\ref{Fig_tracking_cluster_xyz}a and Fig.~\ref{Fig_tracking_cluster_xyz}b respectively. For clarity, typical clusters with lifetimes exceeding 20~UE positions are depicted. The tracked clusters are depicted as dots connected by lines and are labeled according to their corresponding physical object. The results reveal a clear dynamic behavior of the different clusters. For example, clusters associated with sidewall~B exhibit significant variations in the X coordinates of their IO centers as the UE moves, while the Y coordinates remain relatively stable. This suggests that these clusters move along the sidewall with the movement of the UE. On the other hand, clusters associated with desk~C exhibit small changes in both coordinates, indicating that these clusters generally remain static when the UE is moving. The consistency between the tracked clusters and the underlying propagation environment demonstrates the well-matched performance of the proposed tracking framework.

\section{Cluster-level channel characterization}
\subsection{Visible clusters in multi-link channels}
Typically in cluster-based channel models~\cite{L. Liu2012, J. Flordelis2020,J. Li2019}, if a cluster contributes to a channel, it is viewed as `visible' in the channel, which is also defined as an `effective cluster' in 3GPP~TR~38.901~\cite{3gpp2024}. The cluster visibility in multi-link channels at a specific snapshot $n$ is evaluated by defining
\begin{equation}
\begin{split}
V_{c,k}^{(n)}=\begin{cases}1, \text{cluster $c$ is visible in the panel~$k$-UE link} \\0, \text{otherwise} \end{cases}
\end{split}
\end{equation}
and $V_{c,k}^{(n)}=1$ if and only if
\begin{equation}
\frac{{\textstyle \sum_{j}^{j\in\eta_c^{(n)},j\in\varsigma_k^{(n)}}} \left | \alpha_j \right |^2  }{{\textstyle \sum_{j}^{j\in\eta_c^{(n)}}} \left | \alpha_j \right |^2}>\delta_{\text{c,th}}
\label{criteria2-1}
\end{equation}
and
\begin{equation}
\frac{{\textstyle \sum_{j}^{j\in\eta_c^{(n)},j\in\varsigma_k^{(n)}}} \left | \alpha_j \right |^2  }{{\textstyle \sum_{j}^{j\in\varsigma_k^{(n)}}} \left | \alpha_j \right |^2}>\delta_\text{p,th}, 
\label{criteria2-2}
\end{equation}
where $\eta_c^{(n)}$ is the set of MPCs assigned to cluster $c$ at snapshot $n$, and $\varsigma_k^{(n)}$ is the set of MPCs in the panel~$k$-UE link. The power ratio thresholds $\delta_{\text{c,th}}$ and $\delta_\text{p,th}\in [0,1)$. The criteria in~ (\ref{criteria2-1}) ensures that the MPCs of the link contribute with sufficient power to cluster $c$, while the criteria in~(\ref{criteria2-2}) ensures that cluster $c$ constitutes a significant portion of the total power received for the specific link.
Then, the total number of visible clusters in the link to panel~$k$ at snapshot $n$ is given by $N_{\text{c},k}^{(n)}=\sum_{c=1}^{C^{(n)}}V_{c,k}^{(n)}$.

\begin{figure}[t!]
	\centerline{\includegraphics[width=0.49\textwidth]{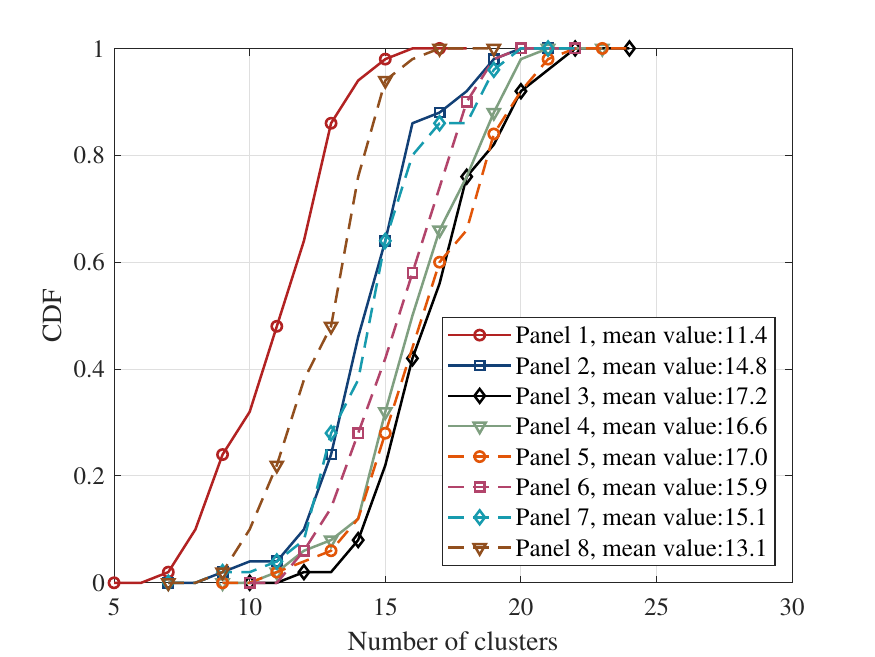}}
	\caption{CDFs of the number of visible clusters in the links to the different panels. }
	\label{Fig_ClusterNum_8panels}
\end{figure}

The cumulative distribution functions (CDFs) of the measured $N_{\text{c},k}^{(n)}$ for all links to the different panels are illustrated in Fig.~\ref{Fig_ClusterNum_8panels}. The result indicates that more clusters are visible in the links to panel~3$\sim$5, i.e., the panels located in the middle. This is because clusters located on one side of the room tend to contribute with less power to links involving panels on the opposite side, as observed in Fig.~\ref{Fig_example clusters}d and Fig.~\ref{Fig_example clusters}e. Consequently, such clusters are often not visible in these links. In contrast, panels situated in the middle are at moderate distances from clusters on both sides of the room, making them more likely to ``see" clusters from both directions.
\subsection{Cluster visibility at the BS and UE sides}
In the COST 2100 channel model~\cite{L. Liu2012} and its extension~\cite{J. Flordelis2020}, cluster visibility is characterized using the concept of visibility regions (VRs) on both the BS and UE side. Each cluster is associated with one or more UE-side and BS-side VRs (UE-VRs and BS-VRs). A cluster is deemed visible in a specific channel link only when the BS and UE are simultaneously located within the corresponding BS-VRs and UE-VRs. Assuming that the visibility of a cluster at the BS and UE sides is independent, the visibility of a cluster $c$ for panel $k$ can be defined as
\begin{equation}
W_c^{\rm{BS}}(k)=\begin{cases} 1, \text{cluster $c$ is visible at the BS side,}\\\quad\text{i.e., panel $k$ is in the BS-VRs of cluster $c$}, \\0, \text{otherwise},\end{cases}
\end{equation}
which fulfills
\begin{equation}
W_c^{\rm{BS}}(k)=\begin{cases} 1, \text{when } V_{c,k}=1,\\
0, \text{when } V_{c,k}=0\text{ and }\sum_{k'=1}^{K}V_{c,k'}>0.\end{cases}
\end{equation}
Note that if $V_{c,k}=0$, the visibility on the BS side cannot be determined. The condition $\sum_{k'=1}^{K}V_{c,k}>0$ ensures that the UE is simultaneously within the UE-VRs of cluster $c$, making the cluster potentially visible.
Similarly, at snapshot $n$, the cluster visibility at the UE side can be described by
\begin{equation}
W_c^{\rm{UE}}(n)=\begin{cases} 1, \text{cluster $c$ is visible at the UE side,}\\\quad\text{i.e., UE is in the UE-VRs of cluster $c$,}\\0, \text{otherwise,}\end{cases}
\end{equation}
which fulfills
\begin{equation}
W_c^{\rm{UE}}(n)=\begin{cases} 1, \text{when } V_{c}^{(n)}=1,\\
0, \text{when } V_{c}^{(n)}=0\text{ and } \sum_{n'=1}^{N}V_{c}^{(n')}>0.\end{cases}
\end{equation}

\begin{figure}[t]
	\centering
	\subfloat[]
	{
		\begin{minipage}[tb]{0.24\textwidth}
			\centering
			\includegraphics[width=1\textwidth]{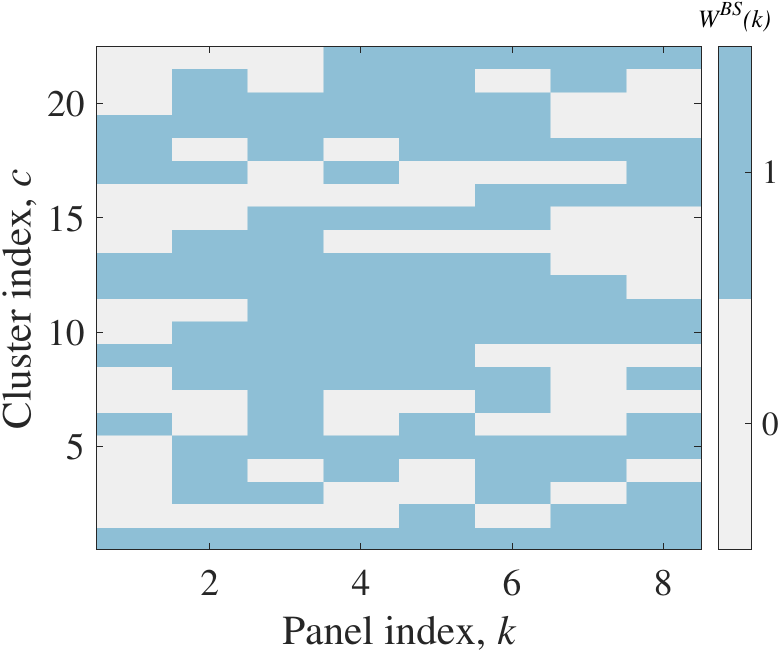}
		\end{minipage}
	}
	\subfloat[]
	{
		\begin{minipage}[tb]{0.24\textwidth}
			\centering
			\includegraphics[width=1\textwidth]{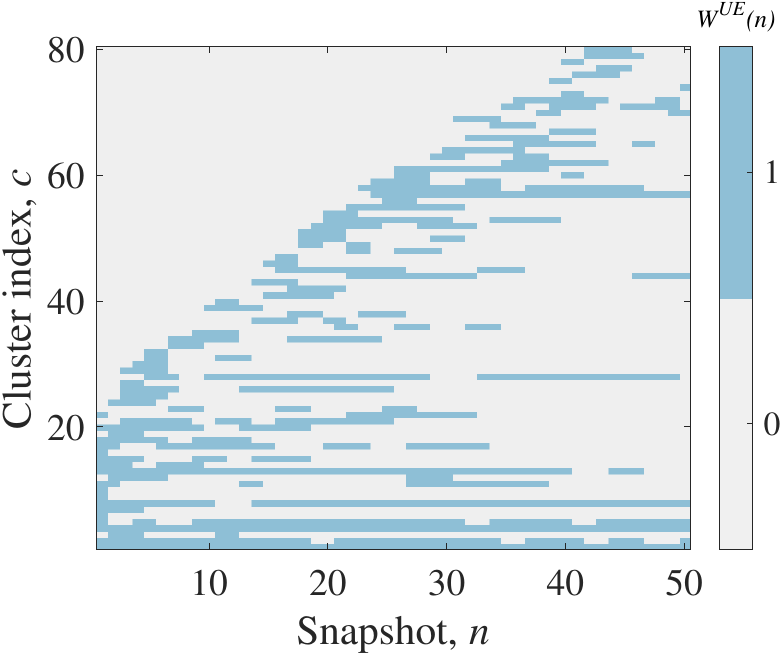}
		\end{minipage}
	}
 \caption{Example of BS cluster visibility $W_c^{\rm{BS}}(k)$ at snapshot $n=1$ and UE cluster visibility $W_c^{\rm{UE}}(n)$ in the link to panel~$1$.}
 \label{Fig_ClusterVisibility}	
\end{figure}

The examples of BS cluster visibility $W_c^{\rm{BS}}(k)$ in snapshot $n=1$ and UE cluster visibility $W_c^{\rm{UE}}(n)$ in the link to panel~$1$ are shown in Fig.~\ref{Fig_ClusterVisibility}a and Fig.~\ref{Fig_ClusterVisibility}b, respectively. The results clearly show the appearance and disappearance of clusters on both the BS and UE side, which correspond to the birth-and-death process of the clusters. This observation highlights that the measured channels exhibit spatial non-wide-sense stationarities (non-WSSs) along the distributed panels and the UE trajectory. In addition, some clusters are observed to temporarily disappear and reappear later across snapshots. According to the COST 2100 channel model~\cite{L. Liu2012}, this is explained by clusters being associated with multiple VRs. As the UE moves, it exits and re-enters different VRs belonging to the same cluster. Similarly, multiple BS-VRs are associated with a single cluster, which accounts for the dynamic visibility of clusters observed across the distributed panels. This dynamic behavior effectively captures the spatial non-WSS and richness of the propagation environment.     

\subsection{Number of VRs per cluster} 
\begin{figure}[t]
	\centering
	\subfloat[]
	{
		\begin{minipage}[tb]{0.24\textwidth}
			\centering
			\includegraphics[width=1\textwidth]{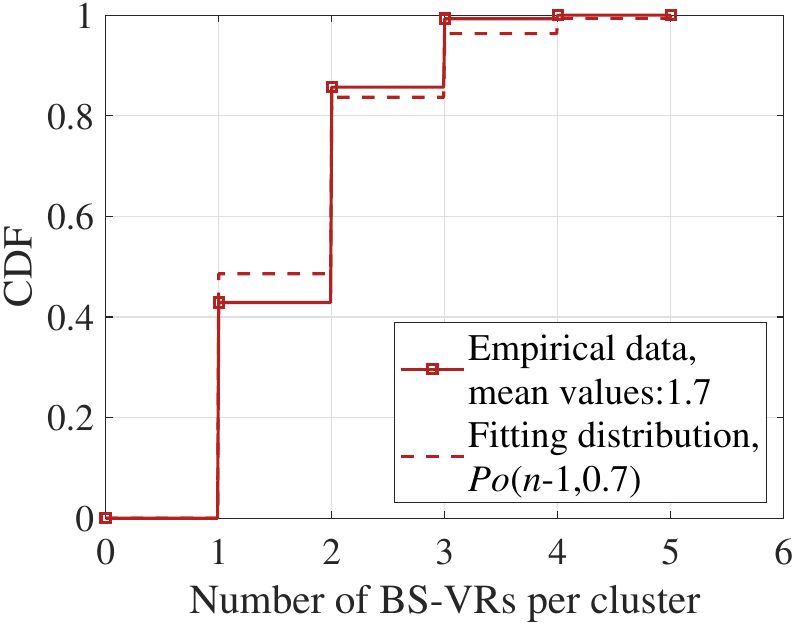}
		\end{minipage}
	}
	\subfloat[]
	{
		\begin{minipage}[tb]{0.24\textwidth}
			\centering
			\includegraphics[width=1\textwidth]{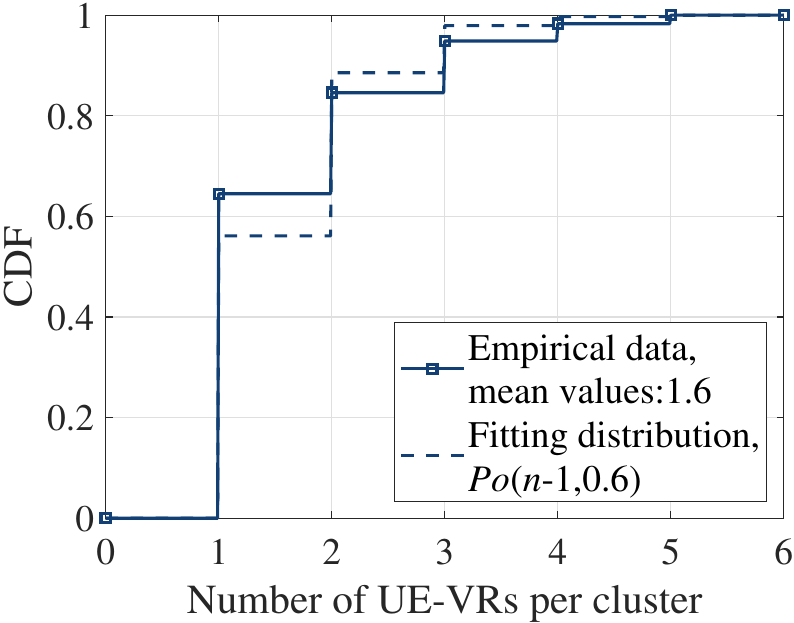}
		\end{minipage}
	}
 \caption{CDFs of the number of (a) BS-VRs and (b) UE-VRs per cluster.}
 	\label{Fig_VRperCluster}
\end{figure}

The number of BS-VRs $N_{\rm{v}}^{\rm{BS}}$ and of UE-VRs $N_{\rm{v}}^{\rm{UE}}$ per cluster are typically modeled as random variables with Poisson distributions~\cite{L. Liu2012,J. Flordelis2020}. Since each cluster is associated with at least one BS-VR and one UE-VR, a 1-shift Poisson distribution is applied here, i.e.,
\begin{equation}
\begin{split}
f_{N_{\rm{v}}^{z}}(n;\lambda_{\rm{v}}^{z} )&=Po(n-1;\lambda_{\rm{v}}^{z})\\
&=\frac{(\lambda_{\rm{v}}^{z})^{n-1}}{\exp(\lambda_{\rm{v}}^{z}-1)(n-1)!}, n\geq 1,     
\end{split}
\end{equation}
where $z\in\{\rm{BS,UE}\}$. The empirical and fitted analytical CDFs of $N_{\rm{v}}^{\rm{BS}}$ and $N_{\rm{v}}^{\rm{UE}}$ are shown in Fig.~\ref{Fig_VRperCluster}a and Fig.~\ref{Fig_VRperCluster}b, respectively. The results indicate that the 1-shift Poisson distribution exhibits a good fit for the empirical CDFs for both $N_{\rm{v}}^{\rm{BS}}$ and $N_{\rm{v}}^{\rm{UE}}$. In addition, the fitted $\lambda_{\rm{v}}^{BS}$ and $\lambda_{\rm{v}}^{UE}$ are found to be similar. To the best of our knowledge, this is the first time that the validity of modeling the number of VRs per cluster as a Poisson-based distribution has been empirically verified using real channel data.
\setcounter{TempEqCnt}{\value{equation}} 
\setcounter{equation}{30}  
\begin{figure*}[hb]
\hrulefill
\begin{equation}
	\begin{split}
 \Lambda(\lambda_Y^{\rm{BS}};L_v^{\rm{BS}}) =& \frac{\lambda^n}{n!} e^{-\lambda (L + \rm{E}(Y > \Delta_0) - 2 \Delta_0) \int_{\Delta_0}^{\infty} f_Y(t; \lambda_Y^{\rm{BS}}) \, dt}
  \left( \prod_{v \in \chi_{00}} f_Y(L_v^{\rm{BS}}; \lambda_Y^{\rm{BS}}) \right)\\ 
& \times \left( \prod_{v \in \chi_{01} \cup \chi_{10}} \int_{L_v^{\rm{BS}}}^{\infty} f_Y(t; \lambda_Y^{\rm{BS}}) \, dt \right)  \left( \int_{L}^{\infty}s(t - L)f_Y(t;\lambda_Y^{\rm{BS}})dt\right)^{ \left | \chi_{11} \right | }   
	\end{split}
	\label{eq:mlfunction}
\end{equation} 
\end{figure*}
\setcounter{equation}{\value{TempEqCnt}}
\subsection{Complete VR length and VR radius}
\begin{figure}[t]
	\centering
	\subfloat[]
	{
		\begin{minipage}[tb]{0.24\textwidth}
			\centering
			\includegraphics[width=1\textwidth]{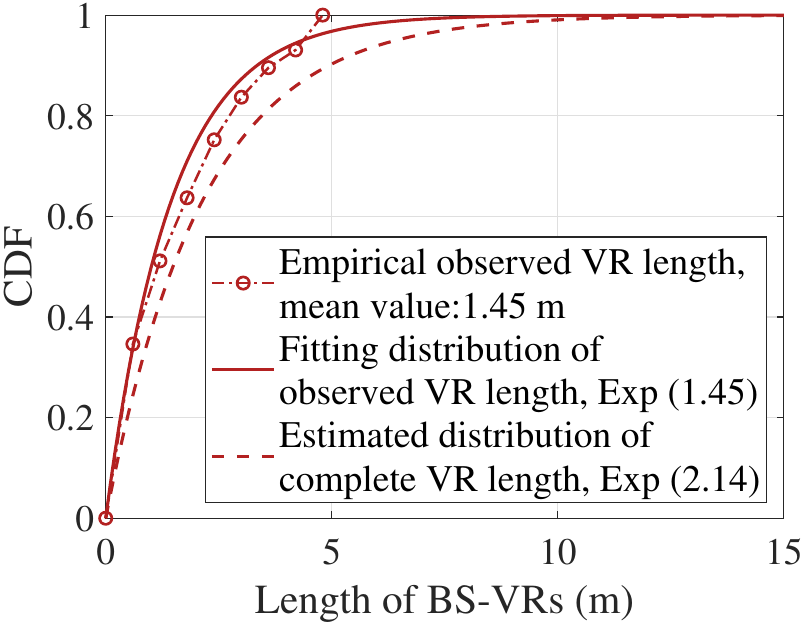}
		\end{minipage}
	}
	\subfloat[]
	{
		\begin{minipage}[tb]{0.24\textwidth}
			\centering
			\includegraphics[width=1\textwidth]{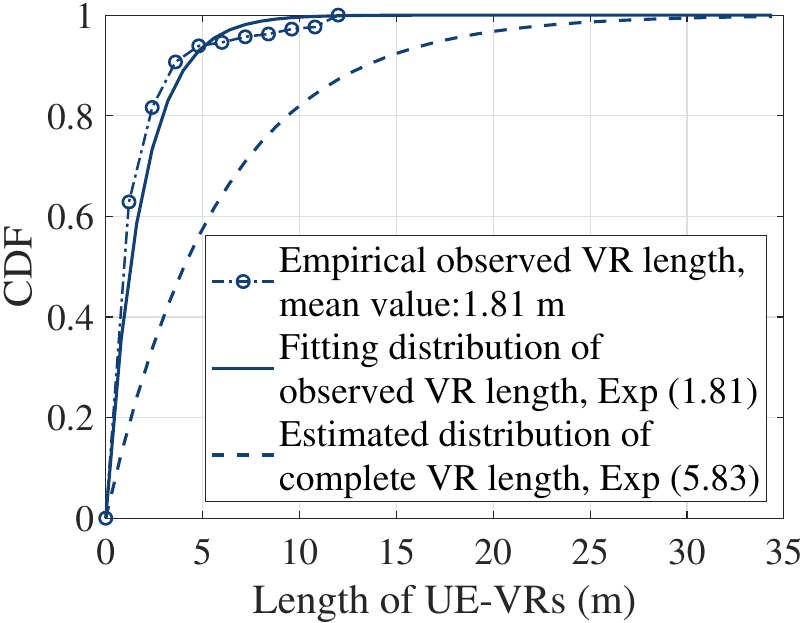}
		\end{minipage}
	}
 \caption{CDFs of the empirically observed lengths, fitting distributions of the observed lengths, and estimated distribution of complete length of the (a) BS-VRs and (b) UE-VRs.}
\label{Fig_VRlength}
\end{figure}

Based on the measured $W_c^{\rm{BS}}(k)$ and $W_c^{\rm{UE}}(n)$, the observed lengths $L_v^{\rm{BS}}$ and $L_{v'}^{\rm{UE}}$ of BS-VR $v$ and UE-VR $v'$ are defined as the maximum lengths along the panels and snapshots in which  $W_c^{\rm{BS}}(k)=1$ and $W_c^{\rm{UE}}(n)=1$, respectively, namely,
\begin{equation}
L_v^{\rm{BS}}=(k_{\text{max},v}-k_{\text{min},v})\cdot\Delta d_{\rm{BS}}
\end{equation}
\begin{equation}
L_{v'}^{\rm{UE}}=(n_{\text{max},{v'}}-n_{\text{min},{v'}})\cdot\Delta d_{\rm{UE}},
\end{equation}
where $k_{\text{max},v}$ ($n_{\text{max},{v'}}$) and $k_{\text{min},{v}}$ ($n_{\text{min},{v'}}$) denote the maximum and minimum panel (snapshot) index fulfilling $\prod_{k'=k_{\text{max},v}}^{k_{\text{min},v}}  W_c^{\rm{BS}}(k')=1$ ($\prod_{n'=n_{\text{max},{v'}}}^{n_{\text{min},{v'}}}  W_c^{\rm{UE}}(n')=1$), and $\Delta d_{\rm{BS}}$ and $\Delta d_{\rm{UE}}$ represent the spacing between neighboring panels and the UE movement distance between neighboring snapshots, respectively. Fig.~\ref{Fig_VRlength}a and Fig.~\ref{Fig_VRlength}b show the empirical and fitted analytical CDFs of $L_v^{\rm{BS}}$ and $L_v^{\rm{UE}}$, respectively. It is found that both the empirically determined $L_v^{\rm{BS}}$ and $L_v^{\rm{UE}}$ can be well fitted using the exponential distributions with the parameters $\lambda_L^{BS}=1.45$ and $\lambda_L^{UE}=1.81$, respectively. 
\begin{figure}[t]
	\centerline{\includegraphics[width=0.40\textwidth]{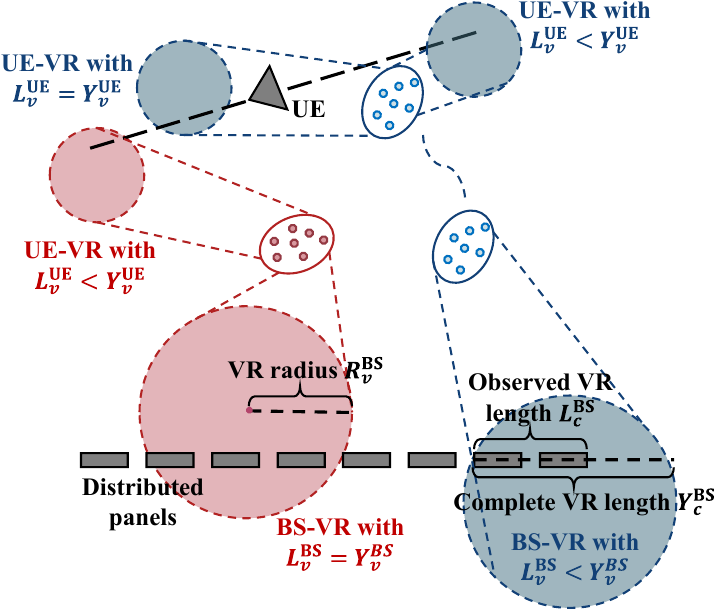}}
	\caption{Relationship between observed and complete length of BS/UE-VRs.}
	\label{Fig_ObservedVRandTrueVR}
\end{figure}

Since the distributed panels and measurement route are always finite, there are clusters whose birth-and-death processes cannot be completely observed~\cite{J. Flordelis2020}. For example, consider the case of BS-VRs: If a cluster is observed to be visible only from the first panel, its complete BS-VR length is likely larger than this observed one, as shown in Fig.~\ref{Fig_ObservedVRandTrueVR}. Similarly, if a cluster is visible for the last panel, its full BS-VR length may also not be fully captured. Therefore, it is inadequate to consider the observed data $L_v^{\rm{BS}}$ as the complete length $Y_v^{\rm{BS}}$ of a BS-VR $v$. To address this, a maximum-likelihood estimator of $Y_v^{\rm{BS}}$ based on $L_v^{\rm{BS}}$ is used.
Firstly, define four sets, i.e.,
\begin{equation}
\begin{split}
    \chi _{00}=\{{\text{VR }v\in\chi_{\text{all}}}:k_{\rm{start}}<k_{\text{min},v},k_{\text{max},v}<k_{\rm{end}}\}\\
\chi _{01}=\{{\text{VR }v\in\chi_{\text{all}}}:k_{\rm{start}}<k_{\text{min},v},k_{\text{max},v}=k_{\rm{end}}\}\\
\chi _{10}=\{{\text{VR }v\in\chi_{\text{all}}}:k_{\rm{start}}=k_{\text{min},v},k_{\text{max},v}<k_{\rm{end}}\}\\
\chi _{11}=\{{\text{VR }v\in\chi_{\text{all}}}:k_{\rm{start}}=k_{\text{min},v},k_{\text{max},v}=k_{\rm{end}}\}
\end{split}
\end{equation}
which consist of BS-VRs that are either observed completely, extended into $k_{\rm{start}}<k_{\text{min},v}$, $k_{\text{max},v}<k_{\rm{end}}$, or into both $k_{\rm{start}}=k_{\text{min},v},k_{\text{max},v}=k_{\rm{end}}$, respectively. The notations $\chi_{\text{all}}$, $k_{\rm{start}}$, and $k_{\rm{end}}$ denote the set including all observed BS-VRs, and the first and last index of the distributed panels, respectively.
Given the good fit shown in Fig.~\ref{Fig_VRlength}a between the empirically determined $L_v^{\rm{BS}}$ and in this case an exponential distribution, $Y_v^{\rm{BS}}$ is modeled as a random variable with the distribution of~\cite{J. Flordelis2020}
\begin{equation}
f_Y(y;\lambda_Y^{\rm{BS}})=\begin{cases} \frac{1}{\lambda_Y^{\rm{BS}}}e^{-\frac{y}{\lambda_Y^{\rm{BS}}}}, y\geq 0, \\0, \text{otherwise.}\end{cases}
\label{distribution of Y_v}
\end{equation}

Given the observed data $L_v^{\rm{BS}}$, the likelihood function of $\lambda_Y^{\rm{BS}}$ is given by~(\ref{eq:mlfunction}),
where $L$, $\Delta_0$, and $\left | \chi_{11} \right |$ represent the distance between the first and last panel, the minimum complete length of BS-VR and the cardinal number of set $\chi_{11}$, respectively. The notation $\lambda$ is the intensity of the total number of BS-VRs in the environment, which is modeled as a Poisson distribution, as suggested in~\cite{L. Liu2012}. For a detailed derivation of~(\ref{eq:mlfunction}), please refer to Appendix~\ref{AppendixA}. By taking the partial derivatives of $\ln(\Lambda(\lambda_Y^{\rm{BS}};L_v^{\rm{BS}}))$ and setting it equal to zero, the estimated $\lambda_Y^{\rm{BS}}$ can be solved and is given~by
\setcounter{equation}{31}
\begin{equation}
\begin{split}
  \hat{\lambda}_Y^{\rm{BS}}=&\frac{\left |(L-\Delta_0)n_0+\Gamma_{\text{BS}})     \right | }{2(n_0-|\chi_{all}|)}\\
  &\cdot\left(1+\sqrt{1+\frac{4(n_0-|\chi_{\text{all}}|)(L-\Delta_0)\Gamma_{\text{BS}}}{((L-\Delta_0)n_0+\Gamma_{\text{BS}})^2}} \right), 
\end{split}
\end{equation}
where $n_0=|\chi_{11}|-|\chi_{00}|$ and $\Gamma_{\text{BS}}=\sum_{v\in\chi_{\text{all}}}(L_v^{\rm{BS}}-\Delta_0)$. The distribution of $Y_v^{\text{BS}}$ is obtained by substituting $\hat{\lambda}_Y^{\rm{BS}}$ into (\ref{distribution of Y_v}).

On the UE side, given the observed UE-VR length $L_v^{\text{UE}}$, the distribution of the complete UE-VR length $Y_v^{\text{UE}}$ can be estimated in the corresponding way. In this case, $\Delta_0$ is set to 0.2~m and 0.24~m at the BS and UE sides, respectively, which approximately correspond to the aperture of the panel and the average distance that the UE moves between two neighboring snapshots.
For comparison, the CDFs of the estimated distributions of $Y_v^{\text{BS}}$ and $Y_v^{\text{UE}}$ are illustrated in Fig.~\ref{Fig_VRlength}a and Fig.~\ref{Fig_VRlength}b, respectively. It is evident from the results that the complete lengths of both BS-VRs and UE-VRs are significantly larger than their observed counterparts. For example, the average complete length (5.83~m) of the UE-VRs is more than three times larger than the average observed length (1.81~m). This highlights the importance of estimating the complete VR length rather than relying solely on observed values.   

Finally, given the estimated distributions of $Y_v^{\text{BS}}$ and $Y_v^{\text{UE}}$ and assuming that VRs can be modeled as circular regions with deterministic radii~\cite{M. Zhu2013}, the radii $R_v^{\text{BS}}$ and $R_v^{\text{UE}}$ of the BS-VR and UE-VR are calculated as $R_v^{\text{z}}=\frac{\pi}{2}\cdot \text{E}\{Y_v^{\text{z}}\}$ where $z\in\{\text{BS, UE}\}$. A detailed derivation of $R_v^{\text{z}}$ can be found in~\cite{M. Zhu2013}. The results of the average observed VR length, the estimated average complete VR length, and the estimated VR radius are summarized in Table~\ref{Table1}.
\begin{table}[t]
\begin{center}
\caption{VR length and radius results}
\label{Table1}
\small
\begin{tabular}{m{1.3cm}<{\centering}| m{1.8cm}<{\centering}| m{1.8cm}<{\centering}| m{1.8cm}<{\centering} }
\hline
\hline
& Observed length (m) & Estimated length (m)& Estimated radius (m)\\
\hline
BS-VRs & 1.45& 2.14&1.36\\
\hline
UE-VRs & 1.81& 5.83&3.71\\
\hline
\hline
\end{tabular}
\end{center}
\vspace{-0.5cm}
\end{table}
\subsection{Common clusters in multi-link channels}
\begin{figure}[t]
	\centerline{\includegraphics[width=0.435\textwidth]{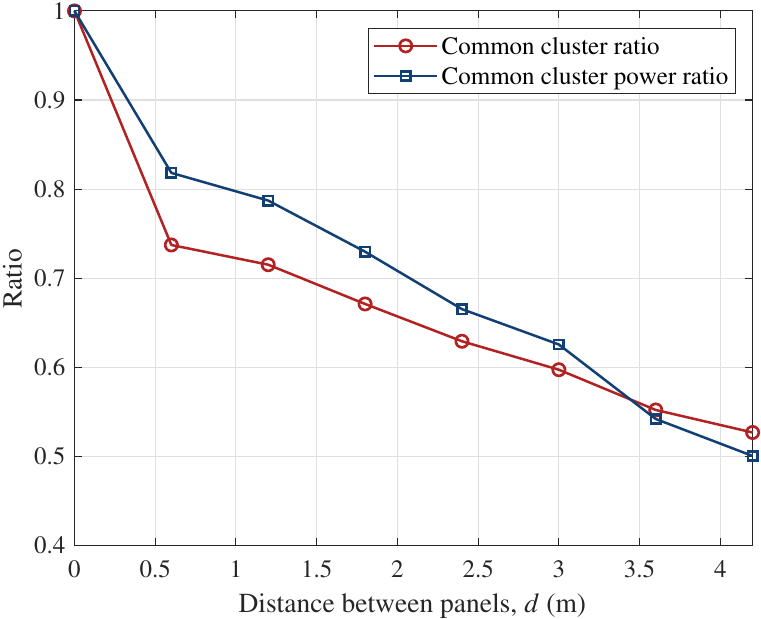}}
	\caption{Measured common cluster ratio $r^{\text{c}}(d)$ and common cluster power ratio $r^{\text{p}}(d)$ with changing distance between panels.}
	\label{Fig_CommonClusterRatio}
\end{figure}

To assess the significance of common clusters in multi-link channels, the dual-link common cluster ratio $r_{k,j}^{\text{c}}$ between the links to panel~$k$ and panel~$j$ is defined as
\begin{equation}
r_{k,j}^{\text{c}}=\mathbf{E}\left\{\frac{|\Gamma_{k,j}^{(n)}|}{|\Gamma_z^{(n)}|}  \right\},
\end{equation}
where $z\in\{k,j\}$, $\Gamma_{k,j}^{(n)} = \Gamma_k^{(n)}\cap\Gamma_j^{(n)}$ represents the set of clusters visible in both links, and $\Gamma_k^{(n)}$ is the set of clusters that are visible in the panel~$k$-UE link at snapshot $n$. Similarly, the common cluster power ratio $r_{k,j}^{\text{p}}$ between links to panel $k$ and $j$ is defined~as
\begin{equation}
\begin{split}
r_{k,j}^{\text{p}}=\mathbf{E}\left\{ \frac{\sum_{c\in \Gamma_{k,j}^{(n)} }\sum_{l\in \eta_c^{(n)}}|\mathbf{A}_l|^2   }{\sum_{c\in \Gamma_z^{(n)} }\sum_{l\in \eta_c^{(n)}}|\mathbf{A}_l|^2   }\right\},
\end{split}
\end{equation}
where $z\in\{k,j\}$. Both the common cluster ratio and common cluster power ratio are further analyzed as functions of the distance between panels for two links, i.e.,
\begin{equation}
r^{\text{c}}(d)=\mathbf{E}\left\{r_{k,j}^\text{c}|_{|k-j|\cdot\Delta d_{BS}=d} \right \}
\end{equation}
\begin{equation}
r^{\text{p}}(d)=\mathbf{E}\left\{r_{k,j}^\text{p}|_{|k-j|\cdot\Delta d_{BS}=d} \right \}.
\end{equation}
The results of $r^{\text{c}}(d)$ and $r^{\text{p}}(d)$ for the measured channels are shown in Fig.~\ref{Fig_CommonClusterRatio}. The results reveal that both ratios decrease as the distance between two panels increases. This is expected because links associated with panels farther apart exhibit more diverse propagation behaviors, resulting in fewer common clusters. It is noteworthy that even for the channels between the most distant panels (e.g., those located in opposite corners of the room), the common cluster ratio remains larger than 0.5. Moreover, the measured $r^{\text{p}}(d)$ indicates that the common clusters contribute to more than half of the total cluster power. These results highlight the significant role of common clusters in measured indoor distributed MIMO channels, demonstrating their importance in multi-link channel analysis.
\subsection{Cluster power, shadowing, and spread}
As suggested in~\cite{M. Zhu2013}, the cluster power is modeled as a function of its delay. Specifically, for a cluster $c$, its power model is expressed as 
\begin{equation}
P_c(\tau_c)=P_0\max\{e^{-k_\tau(\tau_c-\tau_0)},e^{-k_{\tau}(\tau_{\text{cut}}-\tau_0)}\},
\label{cluster_power_model}
\end{equation}
where $P_0$, $k_\tau$, $\tau_c$, $\tau_0$ and $\tau_{\text{cut}}$ are the peak cluster power factor, power decay factor (in the unit of dB/ns), delay of cluster $c$, delay of the LoS component, and cut-off delay, respectively. 
In this study, the cluster power is calculated for each channel link. Given a cluster $c$ that is visible in the panel~$k$-UE link at snapshot $n$, its measured power is calculated as $P_{c,k}^{(n)}=\sum_{l\in \eta_c^{(n)}}|\mathbf{A}_l|^2$. Then $k_\tau$ in (\ref{cluster_power_model}) can be obtained by linear regression of the measured cluster power versus the cluster delay. 
\begin{figure}[t]
	\centerline{\includegraphics[width=0.45\textwidth]{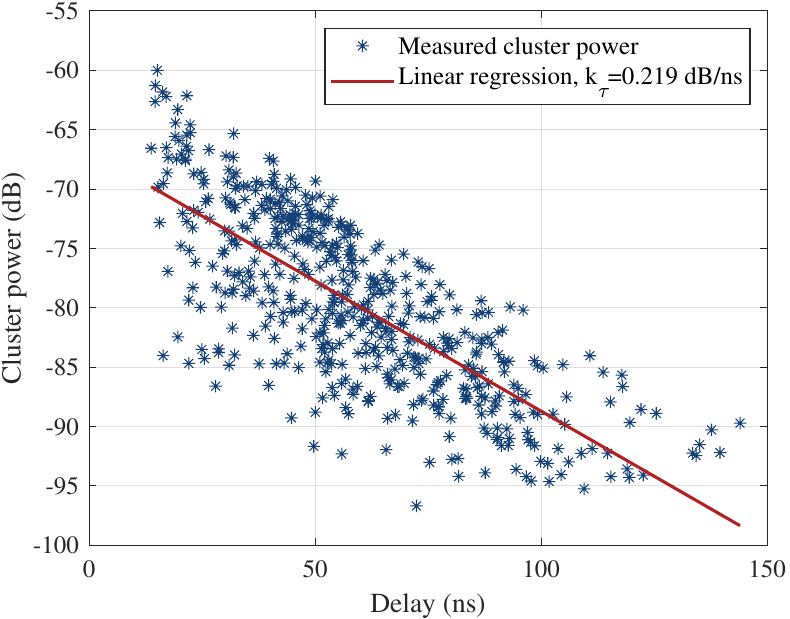}}
	\caption{Measured cluster power and its linear regression in the link with panel~1.}
	\label{Fig_ClusterPower_P1}
\end{figure}

Fig.~\ref{Fig_ClusterPower_P1} illustrates an example of the measured cluster power versus the cluster delay in the link involving panel~1. Through linear regression, the power decay factor is obtained as 0.219~db/ns. The estimated power decay factors for all links are listed in Table~\ref{Table2:Cluster-level_parameters}. It is shown that the power decay factors of the links with panels located in the middle are generally smaller than those of links to panels at the sides. The smallest $k_\tau$ (0.16~dB/ns) is found in the panel~5-UE link while the largest (0.229~dB/ns) is associated with panel~2. By investigating the MPCs in each link, we conjecture that this is because the panels at the sides tend to receive more MPCs from the sidewalls, as shown in Fig.~\ref{Fig_example clusters}d and Fig.~\ref{Fig_example clusters}e. These MPCs are more likely to appear/disappear rapidly, resulting in more severe fast fading of the channels and consequently higher cluster power decay factors~\cite{R. He2015}. 

After performing a linear regression of the cluster power, the cluster shadowing $SF_k$ can be determined as the residuals between the measured cluster power and the regression values~\cite{M. Zhu2013}. The standard deviations of the cluster shadowing for the measured links are listed in Table~\ref{Table2:Cluster-level_parameters}. 

The intra-cluster delay spread and angle spreads are key parameters that reflect the compactness of a cluster and influence the cluster size in the channel model. The intra-cluster delay spread of a cluster~$c$ in the link to panel~$k$ at snapshot~$n$ is defined as 
\begin{equation}
\sigma_{\tau,k}^{(n)}=\sqrt{\left(\frac{\sum_{l\in\eta_c^{(n)}}|\mathbf{A}_l|^2\cdot\tau_{k,l}  }{\sum_{l\in\eta_c^{(n)}}|\mathbf{A}_l|^2}\right)^2- \frac{\sum_{l\in\eta_c^{(n)}}|\mathbf{A}_l|^2\cdot\tau_{k,l}^2  }{\sum_{l\in\eta_c^{(n)}}|\mathbf{A}_l|^2}}. 
\label{delap_spread}
\end{equation}
Similarly, the intra-cluster azimuth and elevation AoA spreads $\sigma_{\phi,k}^{(n)}$ and $\sigma_{\theta,k}^{(n)}$ are calculated using (\ref{delap_spread}) with $\tau_{k,l}$ replaced by the azimuth AoA $\phi_{k,l}$ and elevation AoA $\theta_{k,l}$ of MPC $l$, respectively. The measured $\sigma_{\tau,k}^{(n)}$, $\sigma_{\phi,k}^{(n)}$, and $\sigma_{\theta,k}^{(n)}$ in each link are modeled as log-normal distributions, as suggested in~\cite{L. Liu2012}. The related distribution parameters are summarized in Table~\ref{Table2:Cluster-level_parameters}.

The intra-link cross-correlations among cluster shadowing, delay spread, and angle spreads are evaluated using the cross-correlation coefficients, i.e.,
\begin{equation}
\varpi _k(X,Y)=\frac{\sum_{n=1}^{N}(X_{k}^{(n)} -\bar{X}_k)(Y_{k}^{(n)} -\bar{Y}_k) }{\sqrt{\sum_{n=1}^{N}\left(X_{k}^{(n)} -\bar{X}_k\}\right)^2\sum_{n=1}^{N}\left(Y_{k}^{(n)} -\bar{Y}_k\right)^2} }, 
\end{equation}
where $\bar{X}_k=\mathbf{E}\{X_{k}^{(n)}\}$, $\bar{Y}_k=\mathbf{E}\{Y_{k}^{(n)}\}$ and $X,Y\in\{\sigma_{\tau,k}, \sigma_{\phi,k}, \sigma_{\theta,k},SF_k\}$. The cross-correlations measured in each link are summarized in TABLE~\ref{Table2:Cluster-level_parameters}. From the results, a strong positive correlation is consistently observed between intra-cluster azimuth and elevation angle spreads in all links, with correlation coefficients exceeding 0.7. The cluster shadowing in all links is negatively correlated with the cluster angle spreads, aligning with the findings in~\cite{M. Zhu2011, A. Algans2002}. This indicates that more cluster shadowing is generally associated with larger intra-cluster angle spreads. Additionally, no significant correlation is observed between intra-cluster delay spread and other parameters.  
\begin{table*}[t]
\begin{center}
\caption{Cluster-level statistics of the measured channels}
\label{Table2:Cluster-level_parameters}
\footnotesize
\begin{tabular}{m{0.7cm}<{\centering} m{1.5cm}<{\centering}| m{1.4cm}<{\centering}|m{1.4cm}<{\centering}| m{1.4cm}<{\centering}|m{1.4cm}<{\centering}|m{1.4cm}<{\centering}|m{1.4cm}<{\centering}|m{1.4cm}<{\centering} |m{1.4cm}<{\centering} }
\hline
\hline
& & Panel~1 & Panel~2& Panel~3&Panel~4&Panel~5&Panel~6&Panel~7&Panel~8\\
\hline
\multicolumn{2}{m{2.7cm}<{\centering}|}{ $k_{\tau}$ (dB/$\rm{ns}$)}& 0.219& 0.229&0.193&0.167&0.160&0.175&0.175&0.226\\
\hline
\multicolumn{2}{m{2.7cm}<{\centering}|}{Std. of $SF_{k}$ (dB)}& 4.8& 4.9&5.0&5.2&5.4&4.9&5.1&5.4\\
\hline
\multirow{3}{1.2cm}{Intra-cluster spread} &Fit($\sigma_{\tau,k}$)&$\xi$(0.92,0.25)& $\xi$(0.90,0.28)&$\xi$(0.90,0.29)&$\xi$(0.89,0.31)&$\xi$(0.89,0.28)&$\xi$(0.89,0.29)&$\xi$(0.91,0.32)&$\xi$(0.85,0.33) \\
\cline{2-10}
~&Fit($\sigma_{\phi,k}$)&$\xi$(0.87,0.57)& $\xi$(0.83,0.53)&$\xi$(0.76,0.58)&$\xi$(0.79,0.53)&$\xi$(0.76,0.55)&$\xi$(0.80,0.55)&$\xi$(0.80,0.57)&$\xi$(0.85,0.55)\\
\cline{2-10}
~&Fit($\sigma_{\theta,k}$)&$\xi$(0.56,0.68)& $\xi$(0.45,0.61)&$\xi$(0.39,0.58)&$\xi$(0.38,0.58)&$\xi$(0.33,0.60)&$\xi$(0.40,0.60)&$\xi$(0.39,0.64)&$\xi$(0.54,0.69)\\
\hline
\multirow{6}{0.2cm}{\rotatebox{90}{Correlations}}&$\varpi _k(\sigma_{\tau},\sigma_{\phi})$& 0.08& 0.06&0.08&0.06&0.05&0.06&0&0.1 \\ 
\cline{2-10}
&$\varpi _k(\sigma_{\tau},\sigma_{\theta})$& 0.01& 0.03&0.04&0.02&0&0.01&-0.03&0.1 \\
\cline{2-10}
&$\varpi _k(\sigma_{\tau},SF)$& -0.01& 0.1&0.08&0.14&0.19&0.09&0.13&0.09 \\
\cline{2-10}
&$\varpi _k(\sigma_{\phi},\sigma_{\theta})$& 0.78& 0.81&0.78&0.78&0.78&0.81&0.78&0.77 \\
\cline{2-10}
&$\varpi _k(\sigma_{\phi},SF)$& -0.51& -0.43&-0.37&-0.35&-0.34&-0.32&-0.33&-0.53 \\
\cline{2-10}
&$\varpi _k(\sigma_{\theta},SF)$& -0.50& -0.41&-0.37&-0.33&-0.32&-0.32&-0.35&-0.55 \\
\hline
\hline
\multicolumn{10}{l}{* Fit($X$) represents the fitting log-normal distribution $\xi (\mu_X,\sigma_X)$ of $X$, where $X\in\{\sigma_{\tau,k} (\text{ns}),\sigma_{\phi,k} (^\circ)$,$\sigma_{\theta,k} (^\circ)\}$. Notations $\mu_X$ and $\sigma_X$} \\
\multicolumn{10}{l}{\quad are the mean and variance of the distribution, respectively.}
\end{tabular}
\end{center}
\vspace{-0.5cm}
\end{table*}
\section{Conclusions}
Fully coherent distributed MIMO channels have been measured between eight distributed UPAs and 50 UE positions in an indoor scenario. A novel IO-enabled clustering algorithm has been proposed to jointly cluster the MPCs across multi-link channels, and a Kalman filter-based tracking algorithm has been exploited to dynamically track clusters along with the UE movement. The consistency between the results and the underlying propagation environment demonstrates the performance of the proposed algorithms.
Cluster-level characterization of the measured channels has been investigated, which is important for channel modeling purposes. The key findings are summarized as follows.
\begin{enumerate}
\item 
Clusters exhibit dynamic visibility across different links, with more clusters observed in links involving panels located in the middle of the sidewall. Cluster birth-and-death processes are evident at both link ends.
\item
The validity of modeling the number of VRs per cluster as a Poisson distribution has been empirically verified. The total VR length is estimated through an MLE. The radii of BS-VRs and UE-VRs have been found to be 1.36 and 3.71~m, respectively; these are crucial parameters in cluster-based channel models.
\item 
The common cluster ratio decreases as the distance between panels increases, yet remains above 0.5 even for links with panels with the largest separation. This highlights the substantial role of common clusters in multi-link channels.
\item
A significant positive correlation between intra-cluster azimuth and elevation angular spreads is observed in all links, with correlation coefficients exceeding 0.7. Moreover, cluster shadowing is negatively correlated with cluster angular spreads across all links, while no significant correlation is observed between intra-cluster delay spread and other parameters.
\end{enumerate}
The findings provide new and crucial insights into cluster behavior in multi-link channels, necessary for accurate modeling of indoor distributed MIMO channels. 
\appendices \section{Derivation of~(\ref{eq:mlfunction})}
\label{AppendixA}
For clarity, $\mathcal{L}^{\text{BS}}$ is used to denote the set of observed lengths $L_v^{\text{BS}}, v\in \mathcal{X}_{\text{all}}$. The likelihood function $ \Lambda(\lambda_Y^{\rm{BS}};\mathcal{L}^{\text{BS}})$ of the parameter $\lambda_Y^{\rm{BS}}$ given $\mathcal{L}^{\text{BS}}$  is expressed as~\cite{S. M. Kay1993} 
\begin{equation}
	\begin{split}
 \Lambda(\lambda_Y^{\rm{BS}};\mathcal{L}^{\text{BS}})=&\text{Pr}(n;\mathcal{L}^{\text{BS}})\times \prod_{v \in \chi_{00}}f_{v,\chi_{00}}(\mathcal{L}^{\text{BS}};\lambda_Y^{\rm{BS}})\\
 &\times \prod_{v \in \chi_{01}}f_{v,\chi_{01}}(\mathcal{L}^{\text{BS}};\lambda_Y^{\rm{BS}})\\
 &\times \prod_{v \in \chi_{10}}f_{v,\chi_{10}}(\mathcal{L}^{\text{BS}};\lambda_Y^{\rm{BS}})\\
 &\times \prod_{v \in \chi_{10}}f_{v,\chi_{10}}(\mathcal{L}^{\text{BS}};\lambda_Y^{\rm{BS}}), 
	\end{split}
    \label{eq39}
\end{equation} 
where each term is denoted as \cite{J. Flordelis2020}
\begin{equation}
	\begin{split}
 \text{Pr}(n;\mathcal{L}^{\text{BS}})=\frac{(\lambda K_{\Upsilon })^n}{n!}e^{-\lambda K_{\Upsilon } } 
	\end{split}
    \label{eq40}
\end{equation} 
\begin{equation}
	\begin{split}
f_{v,\chi_{00}}(\mathcal{L}^{\text{BS}};\lambda_Y^{\rm{BS}})=K_{\Upsilon }^{-1}f_Y(\mathcal{L}^{\text{BS}};\lambda_Y^{\rm{BS}})
	\end{split}
\end{equation} 
\begin{equation}
	\begin{split}
f_{v,\chi_{01}}(\mathcal{L}^{\text{BS}};\lambda_Y^{\rm{BS}})&=f_{v,\chi_{10}}(\mathcal{L}^{\text{BS}};\lambda_Y^{\rm{BS}})\\&=K_{\Upsilon }^{-1}\int_{\mathcal{L}^{\text{BS}}}^{\infty }  f_Y(y;\lambda_Y^{\rm{BS}})dy
	\end{split}
\end{equation} 
\begin{equation}
	\begin{split}
f_{v,\chi_{11}}(\mathcal{L}^{\text{BS}};\lambda_Y^{\rm{BS}})=K_{\Upsilon }^{-1}\int_{L}^{\infty }  (y-L)f_Y(y;\lambda_Y^{\rm{BS}})dy,
	\end{split}
    \label{eq43}
\end{equation} 
where $K_{\Upsilon }$ is a normalization constant, which can be determined through the law of total probability, i.e. requiring $\int_{-\infty }^{\infty }f_v(\mathcal{L}^{\text{BS}};\lambda_Y^{\rm{BS}})=1$. The resulting $K_{\Upsilon }^{-1}$ is expressed as
\begin{equation}
	\begin{split}
K_{\Upsilon }=&\int_{\Delta_0}^{L}  (L-y)f_Y(y;\lambda_Y^{\rm{BS}})dy\\&+2\int_{\Delta_0}^{L}(\int_{t}^{\infty}f_Y(y;\lambda_Y^{\rm{BS}})dy)dt\\&+\int_{L}^{\infty }  (y-L)f_Y(y;\lambda_Y^{\rm{BS}})dy.
	\end{split}
\end{equation} 
Now using the identities \cite{J. Flordelis2020}
\begin{equation}
	\begin{split}
\int_{\Delta_0}^{L}(\int_{t}^{\infty}f_Y(y;\lambda_Y^{\rm{BS}})dy)dt = &(L-\Delta_0)\text{Pr}(Y>\Delta_0;\lambda_Y^{\rm{BS}})\\
&+\int_{\Delta_0}^{L}  (y-L)f_Y(y;\lambda_Y^{\rm{BS}})dy
    	\end{split}
\end{equation} 
and 
\begin{equation}
	\begin{split}
    \int_{L}^{\infty }  (y-L)f_Y(y;\lambda_Y^{\rm{BS}})dy=&(\text{E}(Y>\Delta_0)-L)\\
    &\cdot \text{Pr}(Y>\Delta_0;\lambda_Y^{\rm{BS}}),
        	\end{split}
\end{equation} 
$K_{\Upsilon }$ is solved as 
\begin{equation}
	\begin{split}
    K_{\Upsilon } = &((L-\Delta_0)+(\text{E}(Y>\Delta_0)-\Delta_0))\text{Pr}(Y>\Delta_0;\lambda_Y^{\rm{BS}}).
            	\end{split}
                \label{eq:k}
\end{equation} 
As a result, (\ref{eq:mlfunction}) can be obtained by inserting (\ref{eq40})--(\ref{eq43}), and (\ref{eq:k}) into (\ref{eq39}).
\ifCLASSOPTIONcaptionsoff
  \newpage
\fi

\end{document}